\documentclass[onecolumn]{emulateapj}
\usepackage{float}
\usepackage{color}

\shorttitle{Relativistic stars with purely toroidal magnetic fields with 
realistic equations of state}
\shortauthors{Kiuchi et al.}

\begin{document}


\title{Equilibrium Configurations of Relativistic stars with purely toroidal magnetic fields : effects of 
realistic equations of state}


\author{Kenta Kiuchi\altaffilmark{1}}
\affil{Department of Physics, Waseda University, 3-4-1 Okubo,
 Shinjuku-ku, Tokyo 169-8555, Japan~}

\author{Kei Kotake\altaffilmark{2}}
\affil{Division of Theoretical Astronomy/Center for Computational Astrophysics, National Astronomical Observatory of Japan, 2-21-1, Osawa, Mitaka, Tokyo, 181-8588, Japan~}

\and

\author{Shijun Yoshida\altaffilmark{3}}
\affil{Astronomical Institute, Tohoku University, Sendai 980-8578, Japan~}


\altaffiltext{1}{kiuchi@gravity.phys.waseda.ac.jp}
\altaffiltext{2}{kkotake@th.nao.ac.jp}
\altaffiltext{3}{yoshida@astr.tohoku.ac.jp}

\begin{abstract}
We investigate equilibrium sequences of relativistic stars containing purely 
toroidal magnetic fields with four kinds of realistic 
equations of state (EOSs) of SLy (Douchin et al.), 
FPS (Pandharipande et al.), Shen (Shen et al.), and LS (Lattimer \& Swesty). 
We numerically construct thousands of equilibrium configurations 
in order to study the effects of the realistic EOSs. 
 Particularly we pay attention to the equilibrium sequences of constant baryon mass
and/or constant magnetic flux, which model evolutions of an isolated neutron
star, losing angular momentum via gravitational waves. 
Important properties 
obtained in this study are summarized as follows ; 
(1) Unlike the polytropic EOS, it is found for the realistic EOSs that
 the maximum masses do not monotonically increase with 
the field strength along the constant magnetic flux
 sequences. 
(2) The dependence of the mass-shedding angular velocity on the EOSs 
is determined from that of the non-magnetized case. The stars with 
Shen(FPS) EOS reach the mass-shedding limit 
at the smallest(largest) angular velocity, while the stars with 
SLy or Lattimer-Swesty EOSs take the moderate values.
(3) For the supramassive sequences, the equilibrium configurations 
are found to be generally oblate for the realistic EOSs 
in sharp contrast to the polytropic stars. 
For FPS(LS) EOS, the parameter region which permits 
the prolately deformed stars is widest(narrowest). For SLy and Shen EOS, it is in medium.
Furthermore, the angular velocities $\Omega_{\rm up}$,  
above which the stars start to spin up as they lose angular momentum, are found to 
depend sharply on the realistic EOSs. Our analysis indicates that the hierarchy of this spin up angular velocity is 
$\Omega_{\rm up,SLy} > \Omega_{\rm up,FPS} > \Omega_{\rm up,LS}>\Omega_{\rm up,Shen}$ 
and this relation holds even if the sequences have strong magnetic fields. 
Our results suggest the relativistic stars containing purely toroidal 
magnetic fields will be a potential source of gravitational waves 
and the EOSs within such stars can be constrained by observing the angular velocity,
 the gravitational wave, and the signature of the spin up.
\end{abstract}


\section{Introduction}\label{sec:Intro}

Neutron stars observed in nature are magnetized with the typical
magnetic field strength $\sim 10^{11}$--$10^{13}$ G \citep{Lyne}.  
For a special class of the neutron stars such as soft gamma-ray repeaters 
(SGRs) and anomalous X-ray pulsars (AXPs), the field
strength is often much larger than the canonical value as $\sim
10^{15}$ G, and these objects are collectively referred as 
magnetars \citep{Lattimer:2006,wod}.  
Although such stars are estimated to be only a subclass 
($\sim 10 \%$) of the canonical neutron stars~\citep{kouve}, 
much attention has been drawn because they pose many
astrophysically exciting but unresolved problems. 

Giant flaring activities observed in the SGRs have given us good opportunities to study 
the coupling of the interior to the magnetospheric structures \citep{Thompson:1995gw,Thompson:1996pe}, 
but we still know little of the relationship between the crustal fraction and the 
subsequent starquakes (see references in \citet{anna,Geppert:2006cp}). 
The origin of the large magnetic field is also a big problem, whether
 descended from the main sequence stars \citep{Ferrario:2007bt} 
or generated at post-collapse in the rapidly rotating neutron star 
\citep{Thompson:1993hn}. 
Assuming large magnetic fields before core-collapse, 
 extensive magnetohydrodynamic (MHD) stellar collapse
simulations have been carried out recently \citep{Kotake:2004,Obergaulinger:2006,Shibata:2006hr,Livne:2007,luc2007,takiwaki1,sawai2008,Kiuchi:2008ss,takiwaki} 
towards the understanding of
 the formation mechanism of magnetars.
Here it is worth
mentioning that the gravitational waves could be a
useful tool to supply us with the information about magnetar interiors 
\citep{Bonazzola:1995rb,Cutler:2002nw}. 
While in a microscopic point of view, 
effects of magnetic fields larger than the so-called QED 
limit of $B_{\rm QED} = 4.4\times
 10^{13}~\rm{G}$, on the EOSs (e.g., \citet{Lattimer:2006}) and 
the radiation processes have been also elaborately investigated (see
 \citet{Harding:2006} for a review). 
For the understanding
of the formation and evolution of the magnetars, the unification of
 these macroscopic and microscopic studies is necessary, albeit 
not an easy task.

In order to investigate those fascinating issues,
the construction of the equilibrium configuration of magnetars 
may be one of the most fundamental problems. Starting from the pioneering 
study by \citet{Chandra:1953},  extensive studies have been done and this research field is now experiencing  
"renaissance"  \citep{Bocquet:1995,Ioka:2003,Ioka:2004,Kiuchi:2007pa,Konno:1999zv,Tomimura:2005,Yoshida:2006a,Yoshida:2006b}, in which different levels of sophistication in the 
treatment of the magnetic field structure, equations of state (EOSs), 
and the general relativity, have been undertaken. Among them, a more sophistication 
is required for the studies employing the Newtonian gravity, 
because the general relativity(GR) should play
 an important role for the equilibrium configurations of compact objects like 
neutron stars. 
As mentioned below, typical densities for neutron stars interior exceed the nuclear density $\sim 10^{14}{\rm g/cm^3}$. 
In such a high density region, the pressure $P$ and rest mass density $\rho_0$ 
becomes comparable, namely $P\sim \rho_0 c^2$ with $c$ being the speed of light. 
In such a regime, the Newtonian gravity is too weak to gravitationally bind the
 neutron stars without the general relativity, that is $GM/c^2R$ becomes 
the orders of magnitude $10^{-1}$ with $G$, $M$ and $R$ being gravitational constant, 
mass and radius of star. 

It is noted here that the equilibrium configurations of relativistic 
stars without magnetic fields have been elaborately studied using the LORENE code 
\citep{bona}. 
Unfortunately, however, the method of constructing 
a fully general relativistic star with arbitrarily magnetic structures 
is still not available. In most previous studies, 
the weak magnetic fields and/or purely {\it poloidal} 
magnetic fields have been assumed. 
 \citet{Bocquet:1995} and \citet{Cardall:2001} have 
treated relativistic stellar models containing purely poloidal magnetic fields. 
\citet{Konno:1999zv} have analyzed similar models using 
a perturbative approach. 
\citet{Ioka:2004} have investigated structures of mixed poloidal-toroidal 
magnetic fields around a spherical star by using a perturbative 
technique. 

As shown in the MHD simulations of core-collapse supernovae 
(see \citet{Kotake:2006} for a review), 
the toroidal magnetic fields can be efficiently amplified due to 
the winding of the initial seed poloidal fields as long as the core rotates 
differentially. 
After core bounce, as a result, the toroidal fields generically dominate over 
the poloidal ones even if there is no toroidal field initially. 
Even for the weakly magnetized star prior to core-collapse, it should be mentioned that 
 the magnetorotational instability \citep{Balbus:1991} could be another effective 
mechanism for generating large toroidal magnetic fields \citep{Akiyama:2002xn}.
As pointed by  \citet{Spruit}, such large magnetic fields generated by those processes 
may be confined inside the neutron star for thousands of years. 
Moreover, recent calculations of stellar evolution suggest that toroidal component 
of magnetic field dominates over poloidal one with the orders of magnitude $10^{-4}$
in the newly born neutron star after supernova explosion~\citep{Heger:2004qp}. 
These outcomes seem to indicate that some neutron stars could have {\it toroidal} fields 
much higher than the poloidal ones, which motivated \citet{Kiuchi:2008ch} 
to study the equilibrium configurations 
with purely toroidal fields. 
In the study, the master equations were 
derived for the first time for obtaining the relativistic rotating stars with the 
 toroidal fields.
On the other hand, it is well known that pure toroidal magnetic field in star is 
unstable~\citep{Tayler: 1973}. 
\citet{Braithwaite:2005ps} have investigated the evolution of magnetic field in 
stellar interior and found ``twisted torus'' field configuration is stable, 
in which the poloidal and toroidal magnetic field have comparable field 
strength. 
Recently, \citet{Kiuchi:2008ss} have confirmed the Tayler 
instability in the neutron stars 
with the general relativistic magnetohydrodynamics simulation (GRMHD) 
under axisymmetric condition.
Important finding of this study is the toroidal magnetic fields settle down to 
the {\it new} equilibrium states with the circular motion in meridian plane. 
This suggests that the toroidal magnetic field may remain in neutron stars 
even if the poloidal magnetic field is much weaker than the toroidal one. 

However there may 
remain a room yet to be sophisticated in \citet{Kiuchi:2008ch}
  that the polytropic EOSs have been used for simplicity. 
In general, the central density of neutron stars is considered to 
be higher than the nuclear saturation density \citep{bp}. Since we still do not have 
an definite answer about the EOS in such a higher regime, 
many kinds of the nuclear EOSs have been proposed (e.g., \citet{Lattimer:2006}) 
depending on the 
descriptions of the nuclear forces and the possible appearance 
of the exotic physics (e.g., \citet{gled}). While the stiffness of the 
polytropic EOS is kept constant globally inside the star, 
it should depend on the density in the realistic EOSs. 
Since the equilibrium configurations are achieved by the subtle and local balance of the 
gravitational force, centrifugal force, the Lorentz force, and the pressure gradient, 
it is a nontrivial problem how the equilibrium configurations change for the realistic 
EOSs. 

These situations motivate us to study the effects of the realistic EOSs 
on the general relativistic stellar equilibrium configurations with 
purely toroidal magnetic fields.
For the purpose, we incorporate the realistic EOSs to the 
numerical scheme by \citet{Kiuchi:2008ch}. Four kinds of EOSs, of 
SLy \citep{douchin}, FPS \citep{pand}, Shen
\citep{shen98}, and Lattimer-Swesty(LS) \citep{LS} are adopted, which are often
employed in the recent MHD studies relevant for magnetars. By so doing, 
we construct thousands of equilibrium stars for a wide range of parameters 
with different kinds of EOSs.
As shown by \citet{trehan1972}, the toroidal 
 field tends to distort a stellar shape prolately because the toroidal field lines behave like a rubber belt, pulling in the matter around 
the magnetic axis. Such a prolate-shaped neutron star is pointed out to 
be optimal gravitational-waves emitters, depending on 
the degree of the prolateness \citep{Cutler:2002nw}. 
 Thus, we pay attention to how the prolateness is affected by the realistic EOSs. 
For making such configurations, maximum magnetic field strength deep inside 
the magnetars can become as high as $10^{18}$ G, which could be generated 
 due to $\alpha - \Omega$ dynamos in the rapidly rotating neutron 
stars \citep{Thompson:1993hn,Thompson:1996pe}. It should be noted that 
 the possibility of such ultra-magnetic fields has not been rejected so far, because
what we can learn from the magnetar's observations by their periods and spin-down rates 
is only their surface fields ($\sim 10^{15}$ G) .
 We construct equilibrium sequences, along which the total baryon rest 
mass and/or the magnetic flux with respect to the meridional cross-section keep constant. 
It has been suggested that the magnetic field will decay by the process 
of the Ohmic decay, ambipolar diffusion, and Hall drift (e.g., \citet{Goldreich}). 
This implies the magnetic flux might not be well conserved during quasi-steady 
evolution of isolated neutron stars. 
However, taking account into these effect to the neutron star's evolution requires 
extensive efforts. Hence, as a first approximation, we explore the evolutionary 
sequences by use of the equilibrium sequences of constant rest mass and magnetic flux. 

This paper is organized as follows. The master equations for obtaining equilibrium 
configurations of rotating stars with purely toroidal magnetic fields are briefly 
reviewed in Sec.~\ref{sec:basic}. The numerical scheme for computing the stellar models 
and EOSs we employ in this study are 
briefly described in Sec.~\ref{sec:numsch}. Sec.~\ref{sec:result} is devoted to 
showing numerical results. Summary and discussion follow in Sec.~\ref{sec:summary}.
In this paper, we use geometrical units with $G=c=1$.

\section{Summary of basic equations}\label{sec:basic}
Master equations for the rotating relativistic stars containing purely 
toroidal magnetic fields have been derived in \citet{Kiuchi:2008ch}. 
Hence, we only give a brief summary for later convenience. 

Assumptions to obtain the equilibrium 
models are summarized as follows ;  
(1) Equilibrium models are stationary and axisymmetric.
(2) The matter source is approximated by a perfect fluid with infinite conductivity. 
(3) There is no meridional flow of the matter. 
(4) The equation of state for the matter is barotropic. Although we employ 
realistic EOSs in this paper, this barotropic condition can be maintained as
explained in subsection \ref{EOS}. 
(5) The magnetic axis and rotation axis are aligned.

Because the circularity condition (see, e.g.  \citet{Wald:1984}) holds under these assumptions, 
the metric can be written, following~\citet{Komatsu:1989} and~\citet{Cook:1992}, in the form,
\begin{eqnarray}
ds^2 = - {\rm e}^{\gamma+\rho} dt^2 + {\rm e}^{2\alpha}( dr^2 + r^2 d\theta^2 ) 
+ {\rm e}^{\gamma-\rho}r^2\sin^2\theta( d\varphi -\omega dt )^2,\label{eq:metric}
\end{eqnarray}

where the metric potentials, $\gamma$, $\rho$, $\alpha$, and $\omega$, 
are functions of $r$ and $\theta$ only. We see that the non-zero component of 
Faraday tensor $F_{\mu\nu}$ in this coordinate is $F_{12}$.  
In case of purely toroidal magnetic field, the Grad-Shafranov equation is known to be
trivial because it is given only in terms of the $\varphi$ component of the
vector
potential (see Kiuchi \& Yoshida (2008) for details). 
Then the toroidal magnetic field can
be determined by the integrability condition
of the equation of motion (see Eq.(3)).
Integrability of the equation of  motion of the matter requires,
\begin{eqnarray}
e^{\gamma-2\alpha}\sin\theta
F_{12} = K(u);~~~u \equiv \rho_0 h e^{2\gamma}r^2\sin^2\theta,
\label{eq:I-con1}
\end{eqnarray}
where $K$ is an arbitrary function of $\rho_0 h e^{2\gamma}r^2\sin^2\theta$.  
The variables $\rho_0$ and $h$ represent 
the rest mass density and relativistic specific enthalpy, respectively. 
Integrating the equation of motion of the matter, we arrive at the 
equation of hydrostatic equilibrium,
\begin{eqnarray}
\ln h + \frac{\rho+\gamma}{2} + \frac{1}{2}\ln(1-v^2)
+ \frac{1}{4\pi} \int \frac{K(u)}{u}\frac{dK}{du}du = C,\label{eq:Ber}
\end{eqnarray}
where $v=(\Omega-\omega)r\sin\theta {\rm e}^{-\rho}$ with $\Omega$ being the 
angular velocity of the matter and $C$ is an integration constant. 

Here, we have further assumed the rigid rotation. 
To compute specific models of the magnetized stars, we need specify 
the function forms of $K$, which determines the distribution of 
the magnetic fields.
According to \citet{Kiuchi:2008ch}, we take the following 
simple form, 
\begin{eqnarray}
K(u) = b u^k,\label{eq:funcK}
\end{eqnarray}
where $b$ and $k$ are constants. 
Regularity of toroidal magnetic fields on the magnetic axis requires that $k \ge 1$.  
If $k\ge 1$, the magnetic fields automatically vanish at the surface of the star.
\citet{Kiuchi:2008ch} have investigated 
how the choice of $k$ affects the magnetic field 
distribution. In this study, we consider the $k=1$ case  
because in the general relativistic MHD simulation, \citet{Kiuchi:2008ss} have found 
that magnetic distribution with $k\ne1$ is unstable against axisymmetric perturbations. 
Note that 
\citet{Tayler: 1973} has suggested that the magnetic field configuration of $k=1$ 
is unstable against non-axisymmetric perturbations. 

\section{Numerical Method and Equation of State}\label{sec:numsch}
\subsection{Numerical method}
To solve the master equations numerically, we employ the Kiuchi-Yoshida 
scheme~\citep{Kiuchi:2008ch},  
which does not care about the function 
form of EOS. Hence, it is straight forward to update our numerical code for incorporating 
the realistic EOSs discussed below.

After obtaining solutions, it is useful to compute global physical 
quantities characterizing 
the equilibrium configurations to clearly understand the properties of 
the sequences of the equilibrium models. In this paper, we compute the following 
quantities: the gravitational mass $M$, the baryon rest mass $M_0$, 
the total angular momentum $J$, the total rotational energy $T$, the total 
magnetic energy $H$, the magnetic flux $\Phi$, 
the gravitational energy $W$ and the mean deformation rate $\bar{e}$, whose 
definitions are explicitly given in \citet{Kiuchi:2008ch}. 
More explicitly, the mean deformation rate $\bar{e}$ is defined as 
\begin{eqnarray}
\bar{e}\equiv \frac{I_{zz}-I_{xx}}{I_{zz}},\label{eq:meandef}
\end{eqnarray}
where $I_{xx}=\pi\int\epsilon r^4 \sin\theta(1+\cos^2\theta)drd\theta$ 
and $I_{zz}=2\pi\int\epsilon r^4\sin^3\theta drd\theta$ with $\epsilon$ 
being the energy density of the matter. 
(g) Circumferential radius $R_{\rm cir}$ is defined as $R_{\rm cir}\equiv {\rm e}^{(\gamma-\rho)/2} r_e$ with 
$r_e$ being the coordinate radius at the stellar equatorial surface. 
For all the models, checking the relativistic virial identities~
\citep{Bonazzola:1994,Gourgoulhon:1994}, 
we  confirm the typical values are orders of magnitude $10^{-4}-10^{-3}$, which are 
acceptable values for present numerical scheme.

\subsection{Equations of State}\label{EOS}

As mentioned in Sec.~\ref{sec:Intro}, equation of state (EOS) is an
important ingredient for determining the equilibrium configurations.
Instead of conducting an extensive study as done before for the studies of 
rotating equilibrium configurations in which a greater 
variety of EOSs were employed, (e.g., \citet{nozawa,morrison} and references therein), 
we adopt here four kinds of EOSs, of 
SLy \citep{douchin}, FPS \citep{pand}, Shen
\citep{shen98}, and Lattimer-Swesty(LS) \citep{LS}, which are often
employed in the recent MHD studies relevant for magnetars.

In the study of cold neutron stars, the $\beta$-equilibrium condition
with respect to beta decays of the form $e^{-} + p \longleftrightarrow n +
\nu_{e}$ and $n \longleftrightarrow p + e^{-} + \bar{\nu}_e$,
can be well validated as for the static properties. Since neutrinos
and antineutrinos escape from the star, their chemical potentials
vanish at zero-temperature $T=0$ with $T$ being the temperature. Thus 
 the $\beta$-equilibrium condition can be expressed as $\mu_{\rm n} = \mu_{\rm e} + \mu_{\rm
p}$, with $\mu_{\rm n}$, $\mu_{\rm e}$, and $\mu_{\rm p}$ being the
chemical potentials of neutron, electron, and proton, respectively.
 With the charge neutrality condition, we can determine
the three independent thermodynamic variables, (for example the
pressure as $P(\rho, Y_e, T)$, with
$Y_e$ being the electron fraction), can only be determined by a single
variable, which we take to be the density, namely $P(\rho, Y_e(\rho))$ \citep{shapiro}, 
noting here that $T=0$
is assumed for the case of the cold neutron stars. 
Thanks to this, we
can use the formalism mentioned in Sec.~\ref{sec:basic} without violating the
barotropic condition of the EOSs. 
It should be noted that \citet{Reisenegger:2008yk} has recently argued 
the breaking of the barotropic condition leads to the stabilization of the 
magnetic equilibria. 
This tells us the importance to model the neutron stars 
by multicomponent fluid, such as neutrons, electrons, and protons \citep{Reisenegger2001}. 
However, taking into account this effect to equilibrium configuration 
is beyond scope of this article. Furthermore, the barotropic condition 
plays a crucial role to obtain the Bernoulli equation Eq.~(\ref{eq:Ber}).
Hence, in this article, we only consider the barotropic case for simplicity. 

At the maximum densities higher than $\sim 2
\rho_{\rm nuc}$, muons can appear, and higher than $\sim 3
\rho_{\rm nuc}$ \citep{wiringa,akmal}, one may take into the possible
appearance of hyperons \citep{gled}.
However, since the muon contribution to pressure at the higher density has
been pointed to be very small \citep{douchin}, and we still do not have detailed
knowledge of the hyperon interactions, we prefer to employ the above
 neutron star matter, namely $e^{-},n, p$, model to higher densities. 
In the following, we shortly summarize features of the
EOSs employed here.

The Lattimer-Swesty
EOS \citep{LS} has been used for many years as a
standard in the research field of
core-collapse supernova explosions and the subsequent neutron star
formations (see references in \citet{sumi,Kotake:2006}); 
it is based on the compressible drop model for nuclei together
with dripped nucleons. The values of nuclear parameters are chosen
according to nuclear mass formulae and other theoretical studies with
the Skyrme interaction.
The Shen EOS \citep{shen98} is a rather modern one and currently often used
in the research field. The Shen EOS is based on the relativistic
mean field theory with a local density approximations, which has
 been constructed to reproduce the experimental data of masses and
radii of stable and unstable nuclei (see references in \citet{shen98}).
FPS are modern version of an earlier microscopic EOS calculations 
by \citet{fp81}, which employs both two body (U14) and three-body
interactions (TNI). In the SLy EOS \citep{douchin}, neutron-excess
dependence is added to the FPS EOS, which is more suitable for the
neutron star interiors. As for the FPS and SLy EOSs, we use the fitting formulae 
presented in \citet{Shibata:2005ss}.

\section{Numerical Results}\label{sec:result}
In constructing one equilibrium sequence, we have three parameters to choose,
namely the central density $\rho_c$, the strength of the magnetic field 
parameter $b$, and the axis ratio $r_p/r_e$. Changing these parameters, we have to
 seek solutions in as wide parameter range as possible to 
study the properties of the equilibrium sequences in detail.
For clarity,  we categorize the computed models into two, 
namely, non-rotating or rotating and discuss them separately, 
as done in \citet{Kiuchi:2008ch}.

To calculate the non-rotating models, $\rho_c$ and $b$ (see Eq.~(\ref{eq:funcK})) have to be 
given with $\Omega=0$, where $\Omega$ is the angular velocity. 
For the rotating models, one need specify the axis ratio $r_p/r_e$ in addition to 
$\rho_c$ and $b$, where $r_p$ and $r_e$ are the polar and the equatorial radii, 
respectively. To explore properties of the relativistic magnetized star models
 with the realistic EOSs, 
we construct thousands of the equilibrium models for each EOS, namely 
for the non-rotating cases,  
$28\times40$ models in the parameter space of $(\rho_c,\ b)$, and 
for the rotating cases, $28\times40\times10$ models in the parameter space of 
$(\rho_c,\ b,\ r_p/r_e)$. 
Throughout this section, we mainly pay attention to the magnetic
field dominant models, i.e., the models with $H>T$, since we are
concerned with the structure of the magnetar. For some magnetars, the
condition of $H>T$ could be realized because the internal magnetic
field strength could reach an order of $10^{18}$ G and the typical spin 
period is several seconds. 
It should be noted that for the very fast spinning stars in which the dynamo process 
will work, the condition of $H<T$ might be realized and such models are important. 
However, recent observations 
support the fossil origin~\citep{Ferrario:2007bt} rather than the dynamo process and 
our main interest is the observed magnetar so far. 
Hence, in this paper, we mainly refer to the models with $H>T$.

\subsection{Non-rotating models}\label{ssec:nonrot}
First let us consider the static configurations for the 
following two reasons. (1) Since the magnetars and the high field neutron stars 
observed so far are all slow rotators, 
the static models could well be approximated to such stars. (2) In the static models, 
one can see purely magnetic effects on the equilibrium properties because there 
is no centrifugal force and all the stellar deformation is attributed 
to the magnetic stress. 

\begin{figure*}
  \begin{center}
  \vspace*{40pt}
    \begin{tabular}{cc}
      \begin{minipage}{0.5\hsize}
      \includegraphics[width=12.5cm]{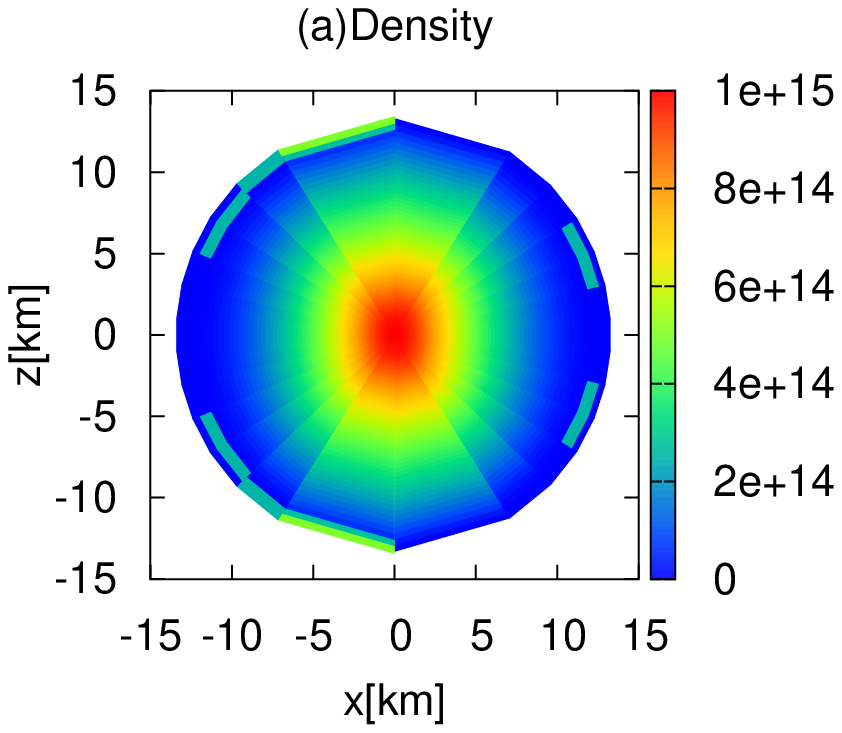}
      \end{minipage}
      \hspace{-1.0cm}
      \begin{minipage}{0.5\hsize}
      \includegraphics[width=12.5cm]{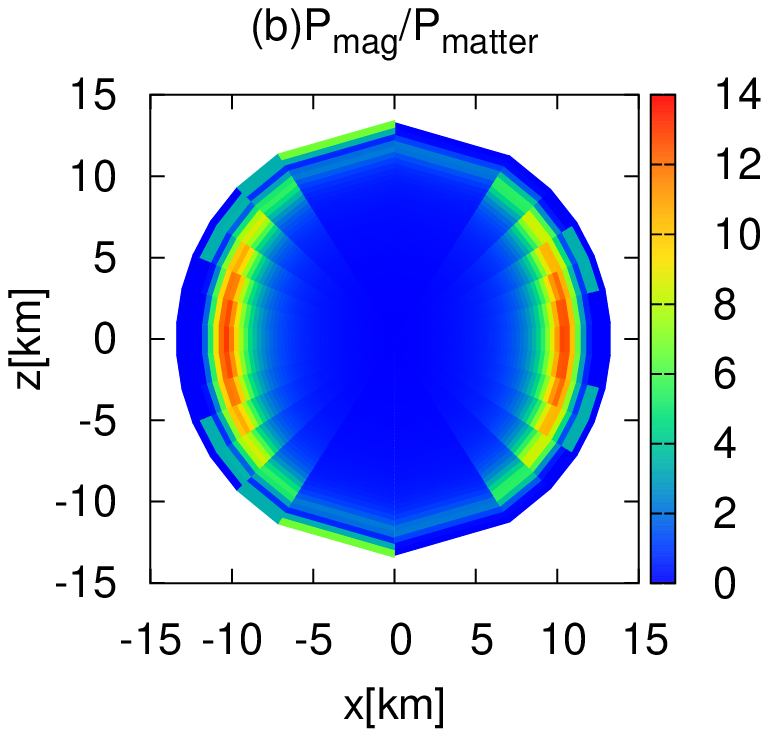}
      \end{minipage}
      \vspace{-0.0cm}
      \\
      \begin{minipage}{0.5\hsize}
      \includegraphics[width=12.5cm]{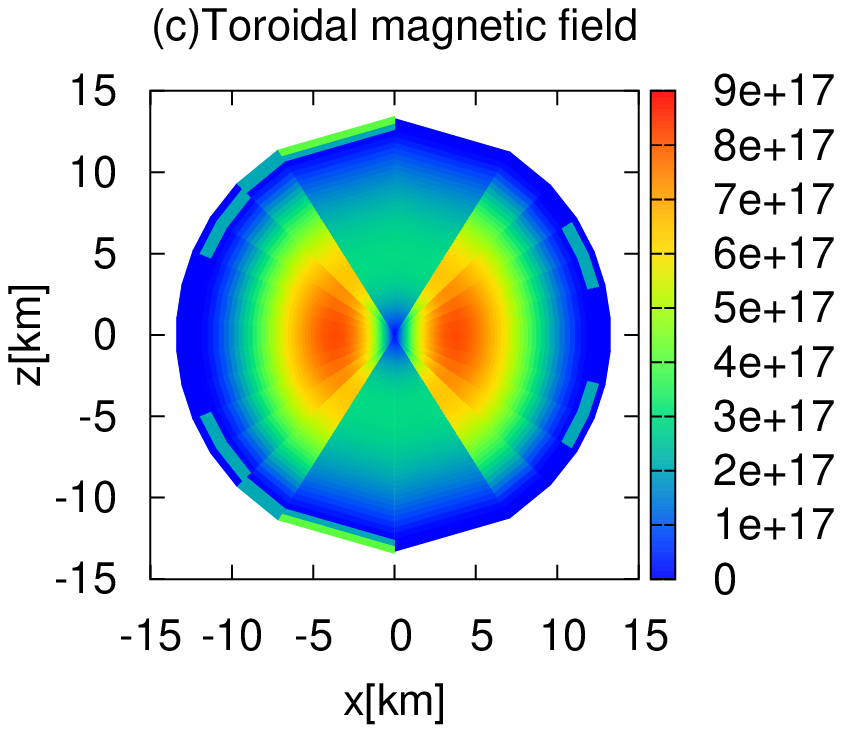}
      \end{minipage}
      \hspace{-0.3cm}
      \begin{minipage}{0.5\hsize}
      \includegraphics[width=10.0cm]{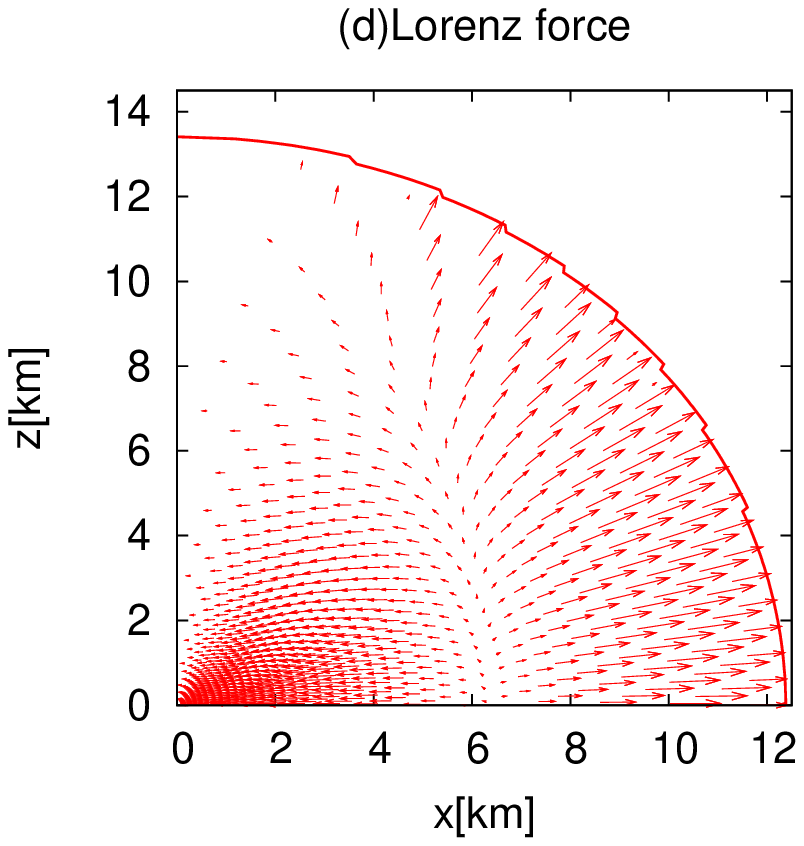}
      \end{minipage}
      \\
      \begin{minipage}{0.5\hsize}
      \includegraphics[width=12.5cm]{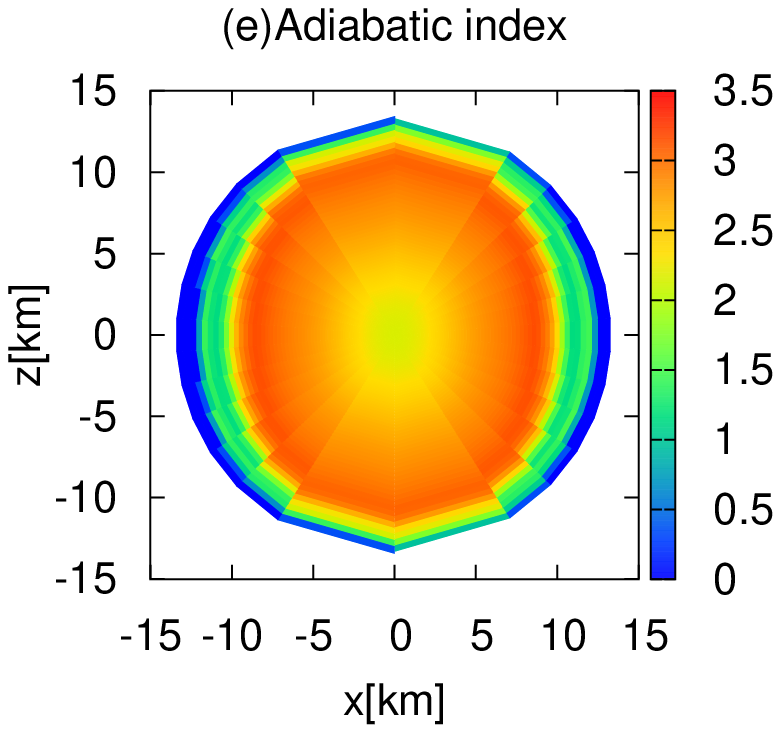}
      \end{minipage}
    \end{tabular}
    \caption{\label{fig:non-rot}Distribution of the rest mass density, 
    the magnetic field, the magnetic pressure ratio to matter pressure, 
    the Lorenz force, and the adiabatic index for the non-rotating star with SLy EOS.
    (a) Equi-density contours, (b) equi-$P_{\rm mag}/P_{\rm matter}$ 
    contours, (c) equi-$B_{(\phi)}$ contours, (d) the Lorenz force, and (e) adiabatic index  
    on the meridional plane for the $M=1.41M_\odot$, 
    $B_{\rm max}=7.8\times 10^{17}$G, $R_{\rm cir}=14.4$[km], 
    and $H/|W|=0.187$ model. In the panel (a)-(c), the solid lines 
    correspond to the stellar surface. In the panel of (d), 
    the thick lines represents the surface.
    }
  \end{center}
\end{figure*}

In Fig. \ref{fig:non-rot}, characteristic quantities 
are given for a $M=1.41M_\odot$ star with the SLy EOS 
characterized by $\rho_c=10^{15}[{\rm g/cm^3}]$, $R_{\rm cir}=14.4{\rm km}$, 
$r_p/r_e=1.14$, $\bar{e}=-0.75$, $H/|W|=0.187$. 
In the figure, it can be seen that the strong toroidal magnetic 
fields peaking in the vicinity of the equatorial plane (panel (c)), 
acts through the Lorentz forces (see the arrows pointing perpendicular to the 
magnetic axis in panel (d)) to pinch the matter around the magnetic axis, 
making the stellar shape prolate (panel (a)). 
The maximum field strength reaches to  $7.8\times10^{17}{\rm G}$ for this model. 
From panel (b), the regions where the magnetic pressure dominates over the matter 
pressure  are seen to be formed. Indeed, the toroidal magnetic field lines behave 
like a rubber belt that is wrapped around the waist of the star. The matter fastened 
by this belt becomes stiff as seen from the large adiabatic indices shown in panel (e), 
where the definition of the adiabatic index is $\partial \ln P/\partial \ln\rho_0|_{s}$ with $s$ being the entropy of the matter ($s=0$ for the cold neutron stars considered here). 
These gross properties are common to the other realistic EOSs.
In the following, we move on to look the differences more in detail.

\begin{figure*}
  \begin{center}
  \vspace*{40pt}
    \begin{tabular}{cc}
      \begin{minipage}{0.5\hsize}
      \includegraphics[width=12.5cm]{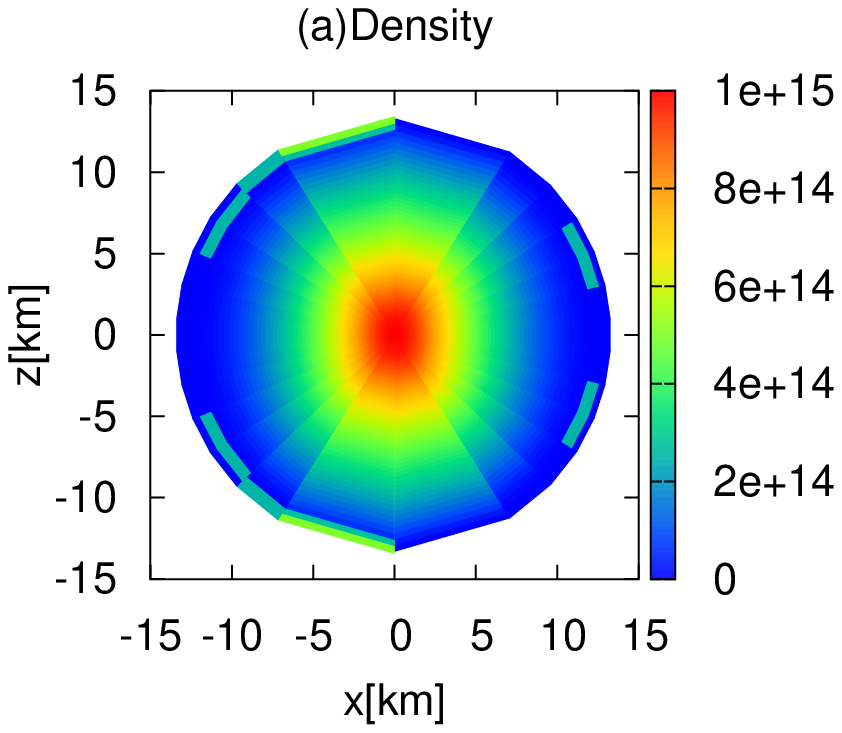}
      \end{minipage}
      \hspace{-1.0cm}
      \begin{minipage}{0.5\hsize}
      \includegraphics[width=12.5cm]{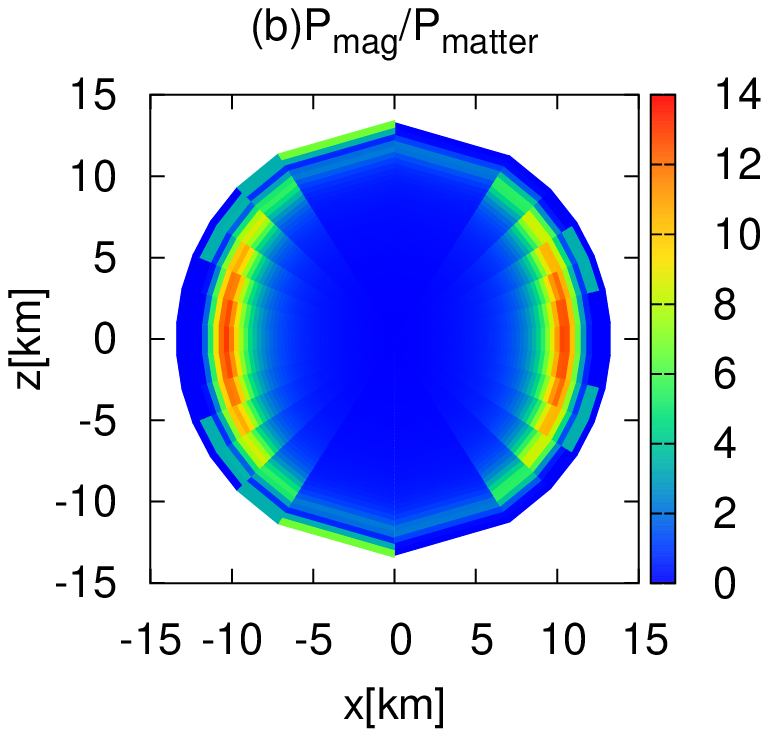}
      \end{minipage}
      \vspace{-0.0cm}
      \\
      \begin{minipage}{0.5\hsize}
      \includegraphics[width=12.5cm]{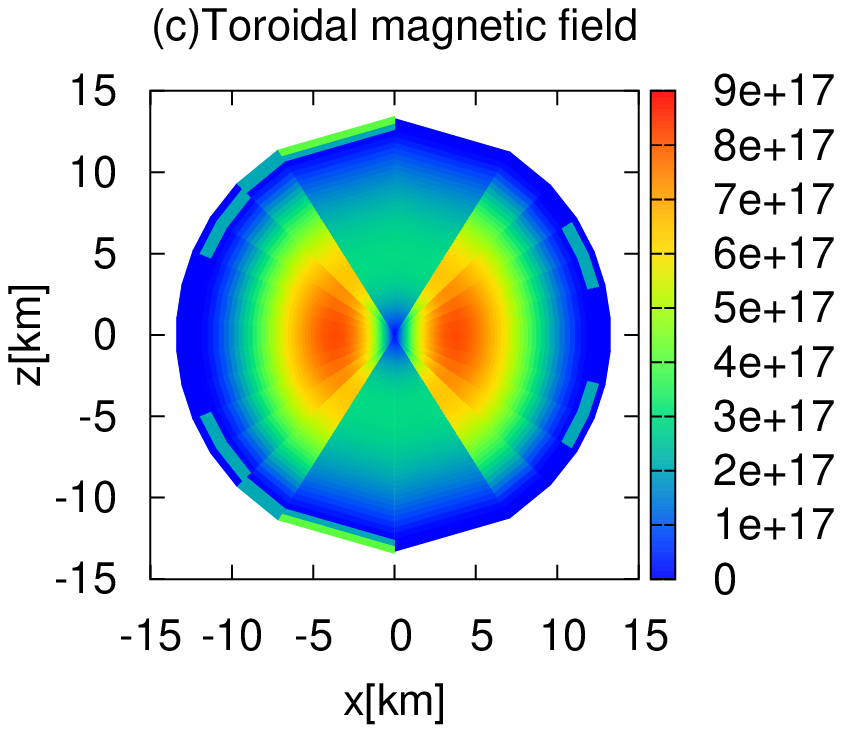}
      \end{minipage}
      \hspace{-0.3cm}
      \begin{minipage}{0.5\hsize}
      \includegraphics[width=10.0cm]{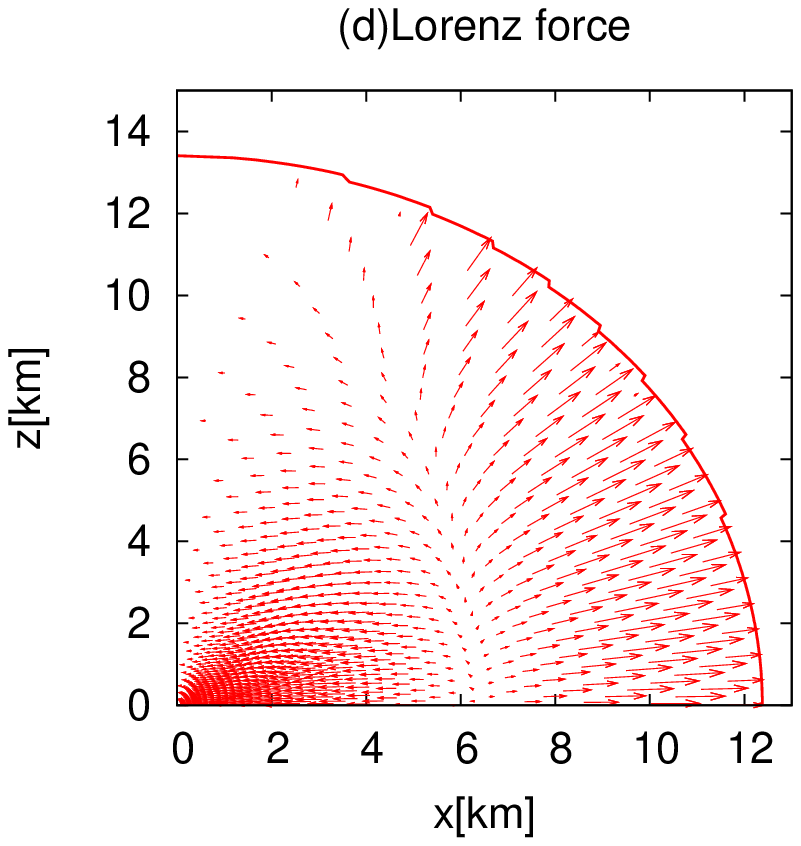}
      \end{minipage}
      \\
      \begin{minipage}{0.5\hsize}
      \includegraphics[width=12.5cm]{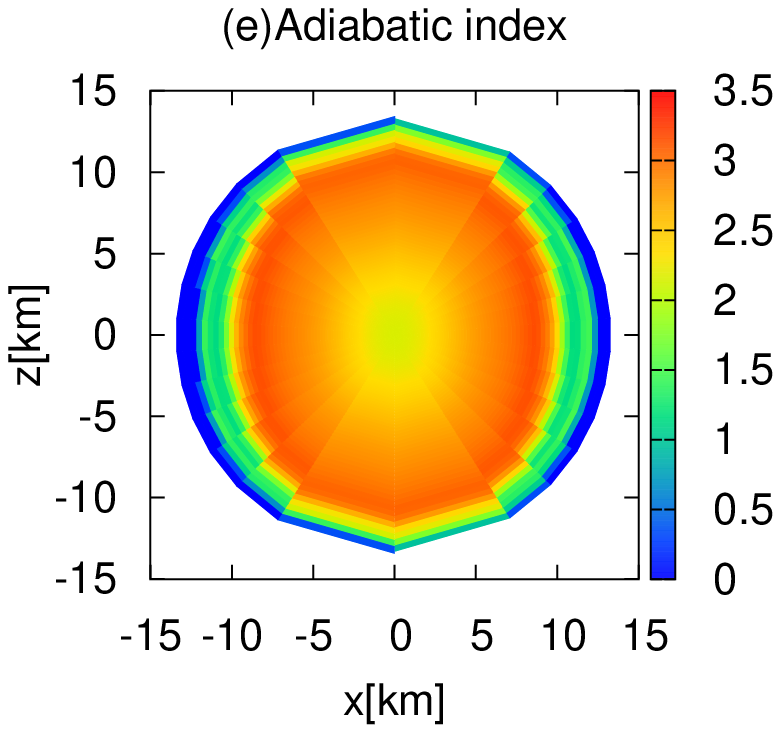}
      \end{minipage}
    \end{tabular}
    \caption{\label{fig:non-rot-FPS}Same as Fig.~\ref{fig:non-rot}, but 
    for FPS EOS, the $M=1.1M_\odot$, 
    $B_{\rm max}=6.9\times 10^{17}$G, $R_{\rm cir}=14.2$[km], 
    and $H/|W|=0.193$ model.
    }
  \end{center}
\end{figure*}

\begin{figure*}
  \begin{center}
  \vspace*{40pt}
    \begin{tabular}{cc}
      \begin{minipage}{0.5\hsize}
      \includegraphics[width=12.5cm]{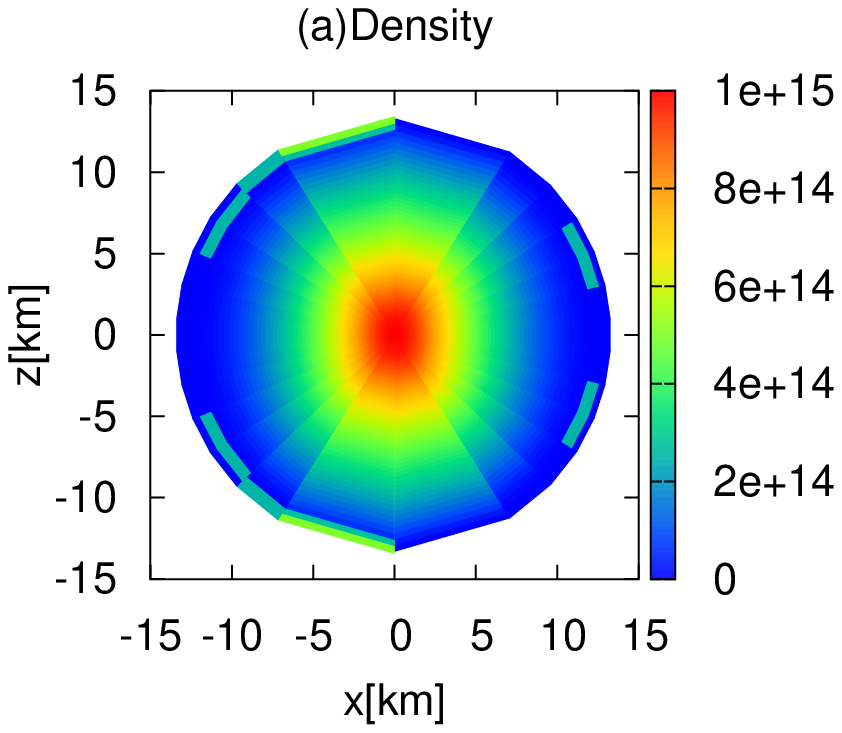}
      \end{minipage}
      \hspace{-1.0cm}
      \begin{minipage}{0.5\hsize}
      \includegraphics[width=12.5cm]{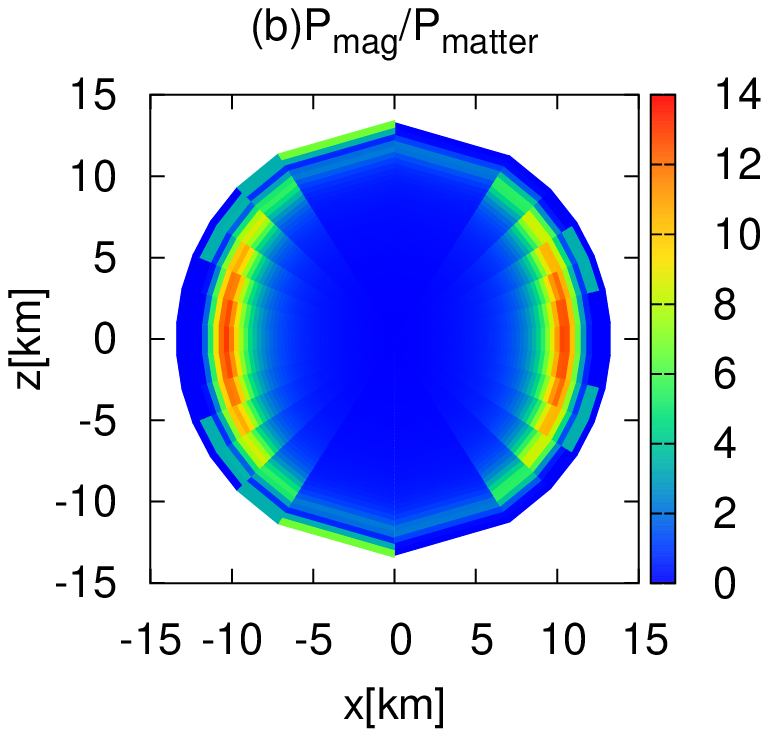}
      \end{minipage}
      \vspace{-0.0cm}
      \\
      \begin{minipage}{0.5\hsize}
      \includegraphics[width=12.5cm]{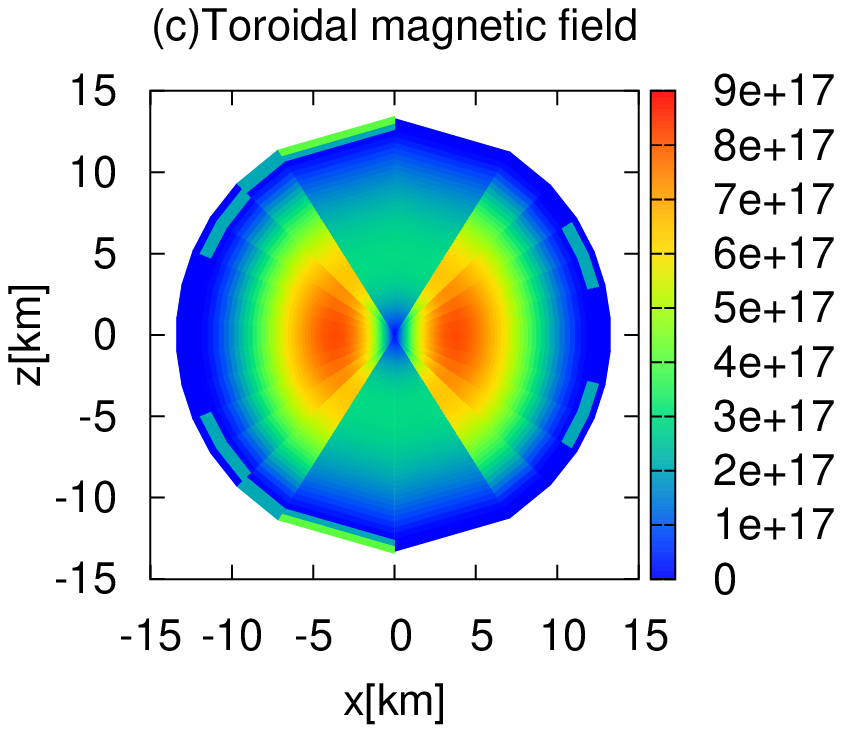}
      \end{minipage}
      \hspace{-0.3cm}
      \begin{minipage}{0.5\hsize}
      \includegraphics[width=10.0cm]{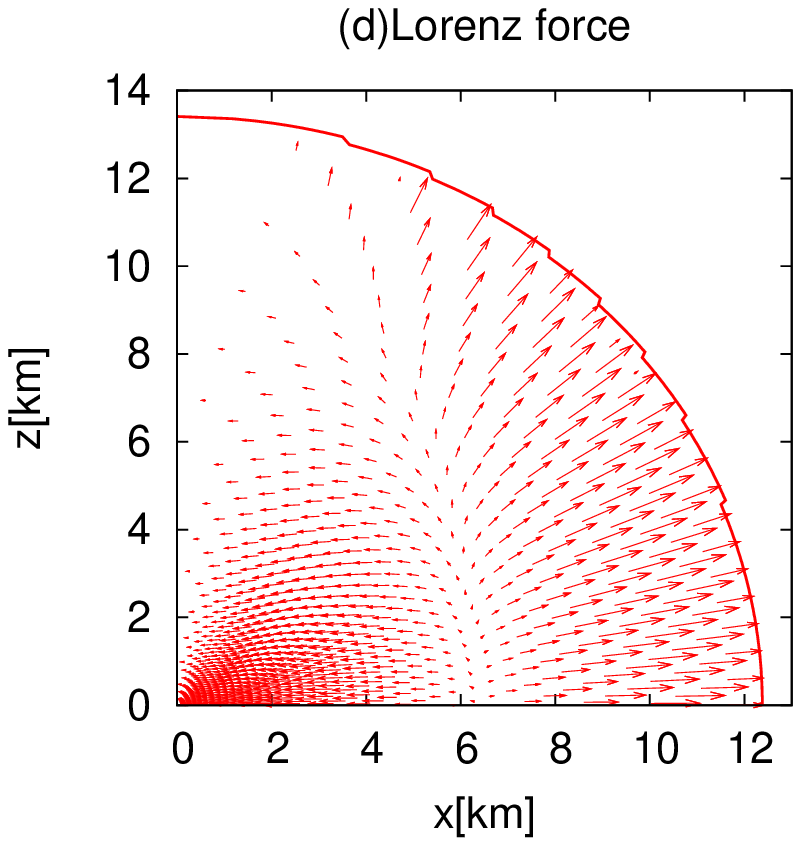}
      \end{minipage}
      \\
      \begin{minipage}{0.5\hsize}
      \includegraphics[width=12.5cm]{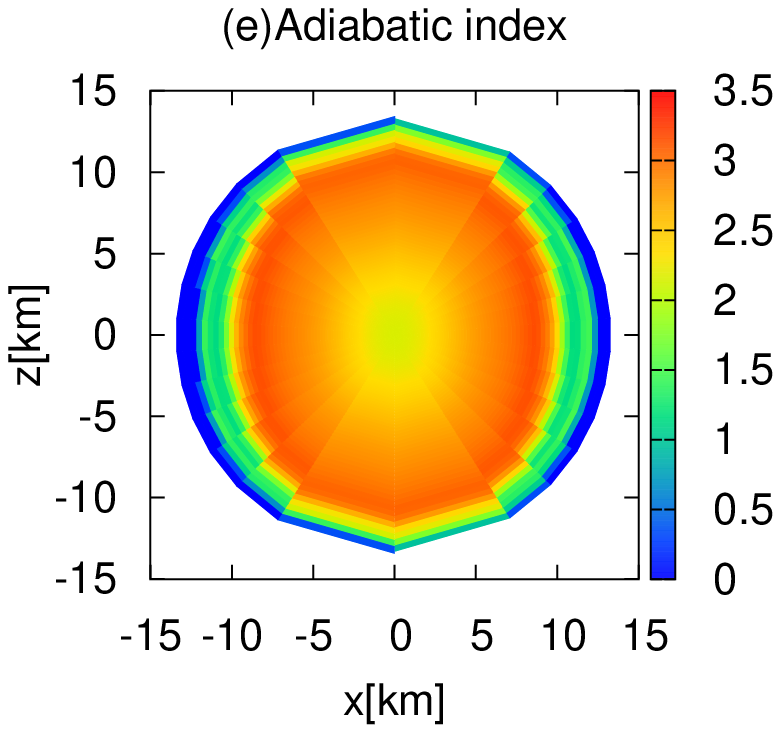}
      \end{minipage}
    \end{tabular}
    \caption{\label{fig:non-rot-Shen}Same as Fig.~\ref{fig:non-rot}, but 
    for Shen EOS, the $M=2.3M_\odot$, 
    $B_{\rm max}=8.2\times 10^{17}$G, $R_{\rm cir}=15.8$[km], 
    and $H/|W|=0.148$ model.
    }
  \end{center}
\end{figure*}

\begin{figure*}
  \begin{center}
  \vspace*{40pt}
    \begin{tabular}{cc}
      \begin{minipage}{0.5\hsize}
      \includegraphics[width=12.5cm]{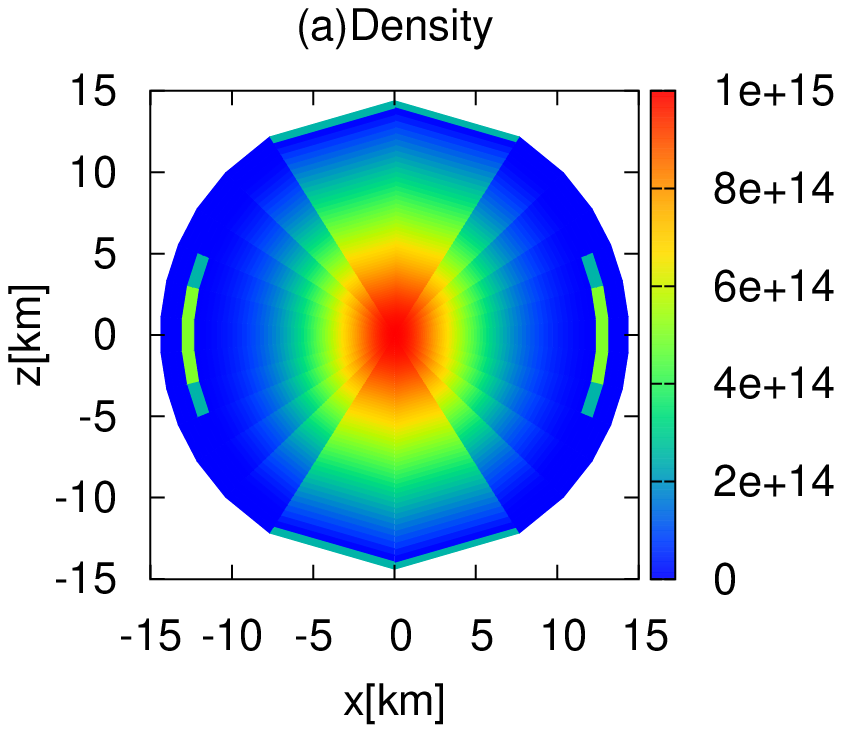}
      \end{minipage}
      \hspace{-1.0cm}
      \begin{minipage}{0.5\hsize}
      \includegraphics[width=12.5cm]{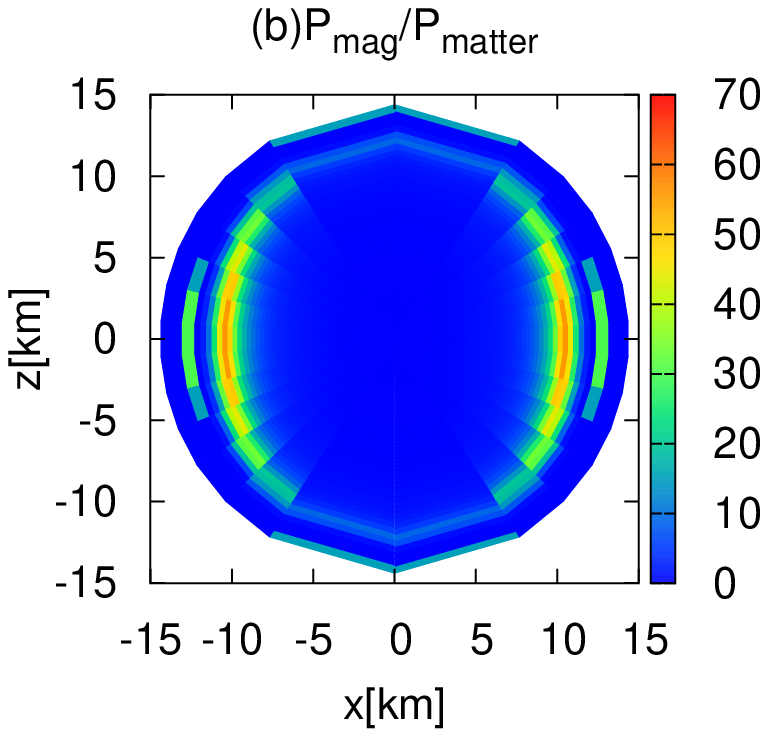}
      \end{minipage}
      \vspace{-0.0cm}
      \\
      \begin{minipage}{0.5\hsize}
      \includegraphics[width=12.5cm]{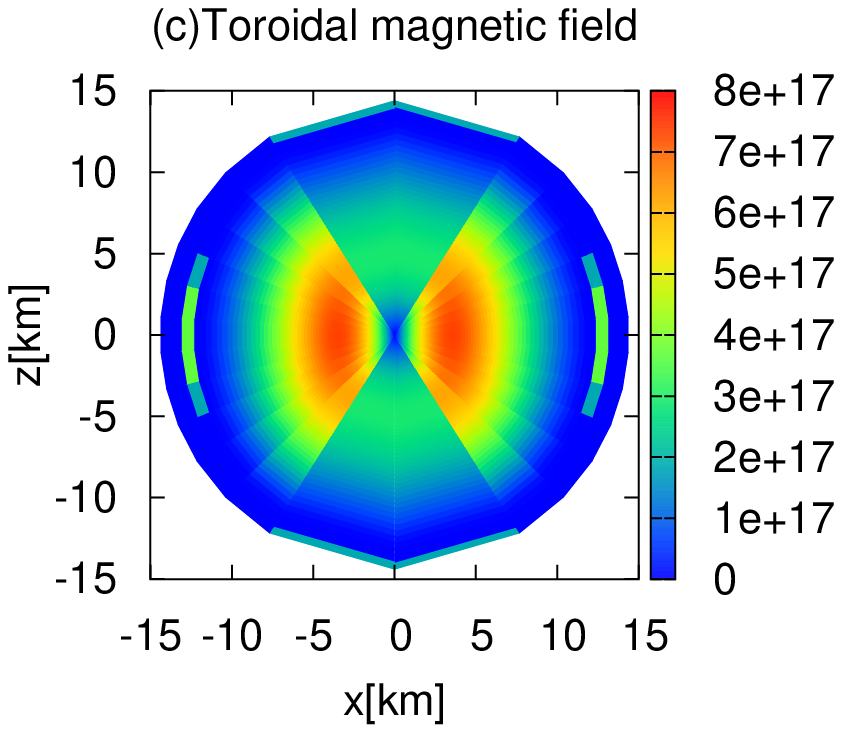}
      \end{minipage}
      \hspace{-0.3cm}
      \begin{minipage}{0.5\hsize}
      \includegraphics[width=10.0cm]{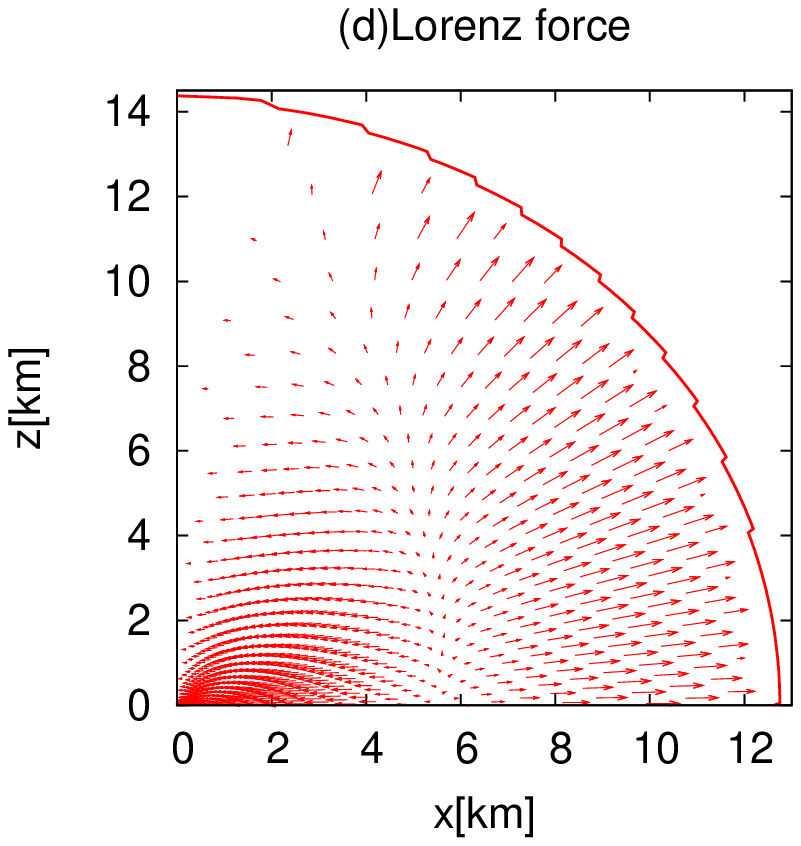}
      \end{minipage}
      \\
      \begin{minipage}{0.5\hsize}
      \includegraphics[width=12.5cm]{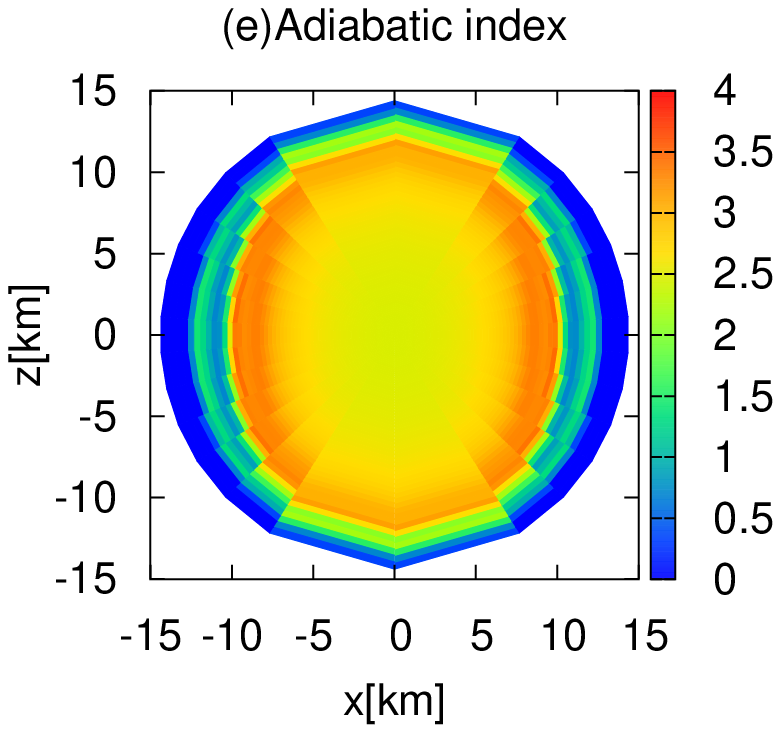}
      \end{minipage}
    \end{tabular}
    \caption{\label{fig:non-rot-LS}Same as Fig.~\ref{fig:non-rot}, but 
    for LS EOS, the $M=1.56M_\odot$, 
    $B_{\rm max}=7.36\times 10^{17}$G, $R_{\rm cir}=14.9$[km], 
    and $H/|W|=0.173$ model.
    }
  \end{center}
\end{figure*}

Figures \ref{fig:non-rot-FPS}-\ref{fig:non-rot-LS} show the equilibrium configurations 
for the other EOSs of FPS, Shen, and LS, plotted in the same style 
as Fig. \ref{fig:non-rot}. The central density for each model is set 
to be the same. Comparing Fig. \ref{fig:non-rot} to Fig.  \ref{fig:non-rot-FPS}, 
 the profiles of the stars with 
SLy and FPS EOSs are quite similar, 
reflecting the resemblance of the stiffness of the 
EOSs below $\rho_c=10^{15}[{\rm g/cm^3}]$ considered here (See Fig. 1 in 
\citet{Kiuchi:2007pa}).
On the other hand, the density distribution of the star with Shen EOS is found to become
less prolate than those with SLy and FPS EOS (compare Fig.~\ref{fig:non-rot}(a) 
with \ref{fig:non-rot-Shen}(a)). 
This can be understood by considering the differences in the distributions of 
the magnetic and matter pressures.
The concentration of the magnetic field to the stellar center for Shen EOS
is weaker than that for SLy or FPS EOS 
(compare Fig.~\ref{fig:non-rot}(c) with \ref{fig:non-rot-Shen}(c)). 
Moreover, the stiffened matter with higher adiabatic indices extends further out for 
Shen EOS than for SLy or FPS EOS, 
(see Fig.~\ref{fig:non-rot}(e) and \ref{fig:non-rot-Shen}(e)), which 
means that the matter pressure stays relatively large up to the stellar surface 
for Shen EOS. As a result, the regions in which the 
magnetic pressure is dominant over the matter pressure can appear rather
in the outer regions for Shen EOS than for SLy or FPS EOS (see Fig.~\ref{fig:non-rot}(b)
-\ref{fig:non-rot-Shen}(c)). This implies that the magnetic fields for Shen EOS  are effectively 
less fastening to pinch the matter around the magnetic axis than those for SLy or 
FPS EOS. For LS EOS, the regions in which the ratio of the magnetic pressure to the matter
 pressure, $P_{\rm mag}/P_{\rm matter}$,
is large also exist near the 
stellar surface (see Fig. \ref{fig:non-rot-LS}(b)). However, 
 the density distribution is found to become similar to that for SLy and FPS EOSs but not for Shen EOS because the magnitude of $P_{\rm mag}/P_{\rm matter}$ is sufficiently higher than that for Shen EOS,
 producing the strong fastening (see the color legend of Fig.~\ref{fig:non-rot-Shen}(b) 
and \ref{fig:non-rot-LS}(b)).

\begin{figure*}
  \begin{center}
  \vspace*{40pt}
    \begin{tabular}{cc}
      \resizebox{80mm}{!}{\includegraphics{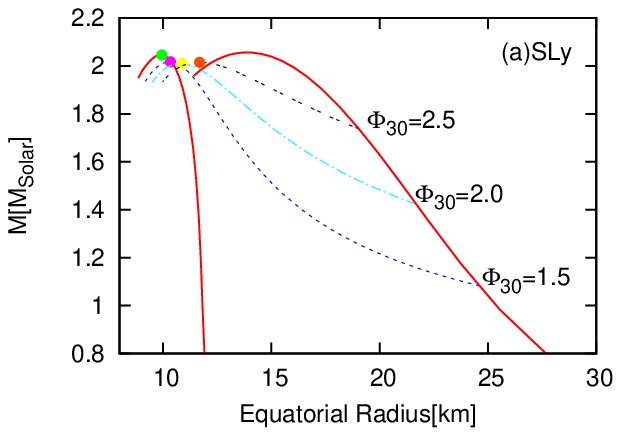}} &
      \resizebox{80mm}{!}{\includegraphics{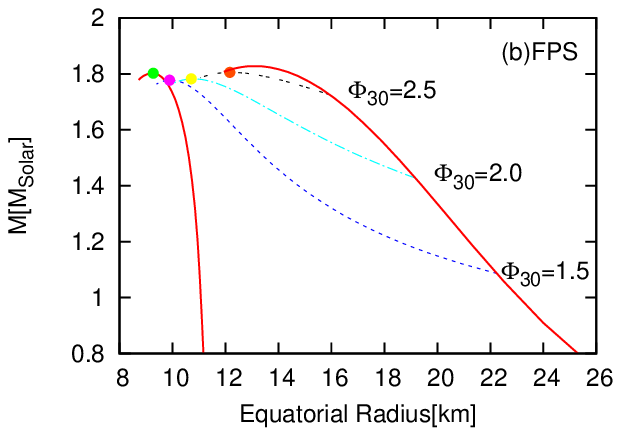}}\\
      \resizebox{80mm}{!}{\includegraphics{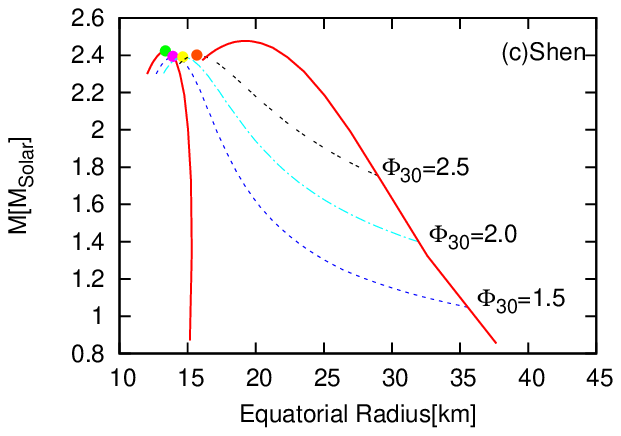}} &
      \resizebox{80mm}{!}{\includegraphics{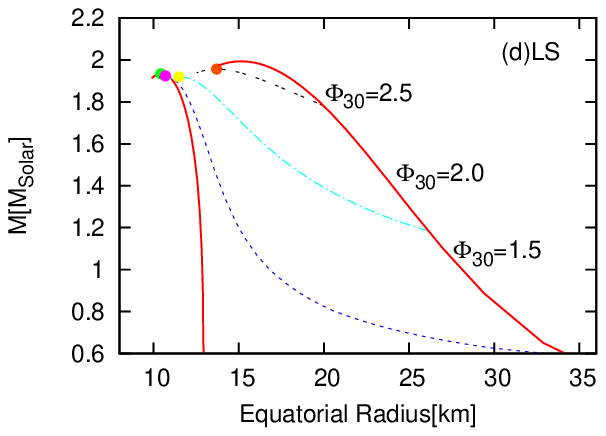}}\\
    \end{tabular}
    \caption{\label{fig:nonrot-qc-ADM}
Circumferential radius versus gravitational mass for models with (a) SLy, (b) FPS, (c) Shen, 
and (d) LS EOS. The left thick line represents spherical, non-magnetized configurations, 
and the right thick line represents the boundary beyond which 
converged solution could not be obtained. Along the dotted lines labeled with $\Phi_{30}$, 
the configurations have constant magnetic flux, and $\Phi_{30}$ denotes the flux value 
normalized by $10^{22}{\rm G~cm^2}$.
The filled circles indicate the maximum gravitational mass models whose  
global physical quantities are given in Table~\ref{tab:nonrot-ADMmax}. 
}
  \end{center}
\end{figure*}

In Fig.~\ref{fig:nonrot-qc-ADM}, the gravitational mass $M$ 
for the different EOSs with the constant magnetic flux sequences 
 is shown as a function of the circumferential radius $R_{\rm cir}$.
 Note that 
$\Phi_{30}$ means the flux normalized by units of $10^{30}{\rm G~cm^2}$. 
and that the curves labeled by their values of $\Phi_{30}$ indicate the sequences  
with the constant magnetic flux.
In each panel, the left thick red line represents the 
spherical star limit and the right thick red line does the non-convergence limit 
($b\sim 10$ in Eq.~(\ref{eq:funcK})) 
beyond which any solution cannot converge with the present numerical scheme 
\citep{Kiuchi:2008ch}. 
Strictly speaking, values of $b$ at the non-convergence limit depend on the central density 
$\rho_c$. However, we confirm the non-convergence limit appears around $b=10$ for 
all the central densities we set in this study. 
  Thus, the non-convergence limits (the right thick red lines in 
Fig.  \ref{fig:nonrot-qc-ADM}) look like smooth curves.
For all the EOSs, the physically acceptable solutions exist in 
the regions bounded by these two lines. The filled circles on each line
represents the maximum mass models. 
To focus on the sequence with the maximum mass, 
which we call as the maximum mass sequences henceforth, in Table~\ref{tab:nonrot-ADMmax}, 
 we summarize their global physical quantities: 
the gravitational mass $M$, the baryon rest mass $M_0$, the circumferential radius 
at the equator $R_{\rm cir}$, the maximum strength of the magnetic fields 
$B_{{\rm max},18}$ normalized by $10^{18}[{\rm G}]$, 
the ratio of the magnetic energy to the gravitational energy 
$H/|W|$, and the mean deformation rate $\bar{e}$.
It should be noted that, irrespective of the EOSs, the sequences with $\Phi_{30}>2.5$ 
encounter the non-convergence limits before it reaches the maximum mass point 
(see also Table~\ref{tab:nonrot-ADMmax}).

\begin{figure*}
  \begin{center}
  \vspace*{40pt}
    \begin{tabular}{cc}
      \resizebox{80mm}{!}{\includegraphics{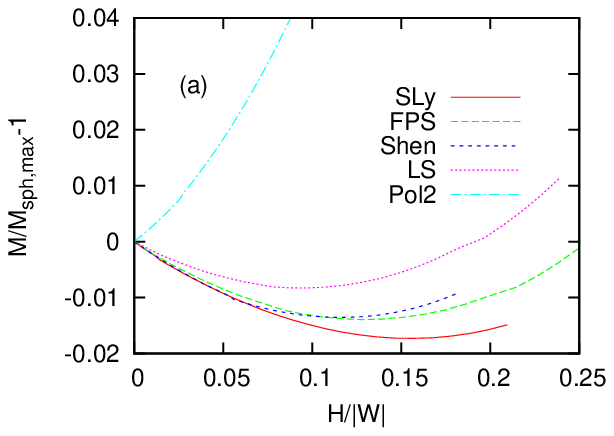}} &
      \resizebox{80mm}{!}{\includegraphics{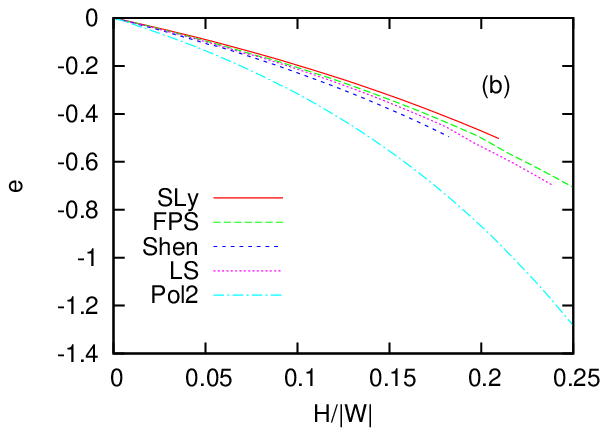}} 
    \end{tabular}
    \caption{\label{fig:Phi-Mmax}
    Dependence of (a) maximum gravitational mass and (b) mean deformation rate 
    on the ratio of the magnetic energy $H$  to the gravitational biding energy $W$. 
    $M_{\rm sph,max}$ denotes the maximum mass of the spherical 
    stars for each realistic EOS. As a reference, the relation under the 
    polytropic EOS is shown as Pol2, where the polytropic index is set to be unity. 
    }
  \end{center}
\end{figure*}

From Table~\ref{tab:nonrot-ADMmax} and Fig. \ref{fig:nonrot-qc-ADM}, 
we can see that regardless of the different EOSs, 
the maximum gravitational masses $M$,
do not increase with the magnetic flux so much, 
in comparison with the other global quantities. 
This is in contrast to the polytropic EOS, in which the maximum masses 
 increase almost monotonically with the magnetic flux \citep{Kiuchi:2008ch}. 
For clarifying the relation, we prepare Fig.~\ref{fig:Phi-Mmax}(a), in which
 the increasing rates of the maximum masses, $(M - M_{\rm spherical})/M_{\rm spherical}$,
 with $M_{\rm spherical}$ being the maximum masses of the spherical stars, are shown
 for the different EOSs along the maximum mass sequences. 
It is seen that the maximum masses, 
irrespective of the different EOSs, 
once decrease, and then begin to increase as $H/|W|$ becomes larger. 
This first decrease is because 
the volume of the high density regions can temporary 
become smaller as the magnetized stars transform to the prolate 
shape due to the magnetic fastening, while keeping the changes in the central density 
small. In fact, the central densities of the maximum mass 
sequences stay almost constant with those of the spherical cases
(see Table~\ref{tab:nonrot-ADMmax}). 
For the polytropic EOS with $n=1$ (referred as Pol2 in Fig. \ref{fig:Phi-Mmax}(a)), 
with $n$ being the polytropic constant, the different trend can be seen. 
The masses keep increasing as the function of $H/|W|$. 
 As shown in Table~\ref{tab:nonrot-ADMmax} (see also Fig. ~\ref{fig:Phi-Mmax}(a)), the maximum gravitational mass 
of the neutron stars with $\Phi_{30} = 2.5$ is larger than that with $\Phi_{30} = 0$ for FPS and LS EOS, 
but smaller for  SLy and Shen EOS.
It should be noted this does not imply the maximum mass with SLy 
or Shen EOS never increases because this is set by the non-convergence 
limit mentioned above.

Fig. \ref{fig:Phi-Mmax}(b) shows the degree of the deformation for the different EOSs
 as a function of $H/|W|$ before the non-convergence limit. 
It is found that the degree can reach to 
an order of $\bar{e}=-O(0.1)$ irrespective of the EOSs. 
Comparing with the polytropic case (Pol2), we find the 
mean deformation rates become smaller for the models with the realistic EOSs. 
The reason is as follows. 
The magnetic belt effects, due to which the matter is pinched, subside in the vicinity 
near the equatorial plane less than $\sim 10$ km (see Fig.~\ref{fig:non-rot}(d)
-\ref{fig:non-rot-LS}(d)). The adiabatic indices there are generally higher than 2 
(see Fig.~\ref{fig:non-rot}(e)-\ref{fig:non-rot-LS}(e)) irrespective of the different 
EOSs. This means that the matter is stiffer than the case of $n=1$ polytropic stars.
 As a result, the magnetic fastening becomes less, leading to the smaller deformation
 for the realistic EOSs.

\begin{table*}
\centering
\caption{\label{tab:nonrot-ADMmax}
Global physical quantities for the maximum gravitational mass models of the constant 
magnetic flux sequences of the non-rotating stars. 
}
\begin{tabular}{cccccccc}
\tableline\tableline
$\Phi_{30}$            &
$\rho_{c,15}$   &
$M[M_{\rm \odot}]$                  &
$M_0[M_{\rm \odot}]$                &
$R_{{\rm cir}}[{\rm km}]$          &
$B_{{\rm max},18}$        &
$H/|W|$                             &
$\bar{e}$                          \\
\tableline
\multicolumn{8}{c}{SLy}\\
\tableline
 0.00E+00& 2.00E+00& 2.05E+00& 2.43E+00& 9.96E+00& 0.00E+00& 0.00E+00&  0.00E+00\\
 1.50E+00& 2.10E+00& 2.02E+00& 2.33E+00& 1.03E+01& 1.16E+00& 8.88E-02& -1.71E-01\\
 2.00E+00& 2.10E+00& 2.01E+00& 2.28E+00& 1.09E+01& 1.47E+00& 1.47E-01& -3.14E-01\\
 2.50E+00& 2.10E+00& 2.02E+00& 2.25E+00& 1.17E+01& 1.73E+00& 2.10E-01& -5.03E-01\\
\tableline
\multicolumn{8}{c}{FPS}\\
\tableline
 0.00E+00& 2.40E+00& 1.80E+00& 2.11E+00& 9.27E+00& 0.00E+00& 0.00E+00&  0.00E+00\\
 1.50E+00& 2.60E+00& 1.78E+00& 2.01E+00& 9.89E+00& 1.40E+00& 1.17E-01& -2.48E-01\\
 2.00E+00& 2.60E+00& 1.78E+00& 1.98E+00& 1.07E+01& 1.73E+00& 1.86E-01& -4.53E-01\\
 2.50E+00& 2.50E+00& 1.81E+00& 1.97E+00& 1.22E+01& 1.93E+00& 2.61E-01& -7.55E-01\\
\tableline
\multicolumn{8}{c}{Shen}\\
\tableline
 0.00E+00& 1.20E+00& 2.42E+00& 2.81E+00& 1.34E+01& 0.00E+00& 0.00E+00&  0.00E+00\\
 1.50E+00& 1.30E+00& 2.39E+00& 2.72E+00& 1.39E+01& 7.26E-01& 7.83E-02& -1.71E-01\\
 2.00E+00& 1.30E+00& 2.39E+00& 2.69E+00& 1.46E+01& 9.09E-01& 1.29E-01& -3.11E-01\\
 2.50E+00& 1.30E+00& 2.40E+00& 2.66E+00& 1.57E+01& 1.06E+00& 1.82E-01& -4.95E-01\\
\tableline
\multicolumn{8}{c}{LS}\\
\tableline
 0.00E+00& 2.00E+00& 1.93E+00& 2.23E+00& 1.04E+01& 0.00E+00& 0.00E+00&  0.00E+00\\
 1.50E+00& 2.10E+00& 1.92E+00& 2.15E+00& 1.12E+01& 1.16E+00& 1.07E-01& -2.31E-01\\
 2.00E+00& 2.10E+00& 1.93E+00& 2.13E+00& 1.21E+01& 1.43E+00& 1.70E-01& -4.18E-01\\
 2.50E+00& 2.00E+00& 1.96E+00& 2.13E+00& 1.37E+01& 1.58E+00& 2.39E-01& -6.98E-01\\
\tableline
\end{tabular}
\end{table*}

\subsection{Rotating models}\label{ssec:rot}
\begin{figure*} [h]
  \begin{center}
  \vspace*{40pt}
    \begin{tabular}{cc}
      \begin{minipage}{0.5\hsize}
      \includegraphics[width=12.5cm]{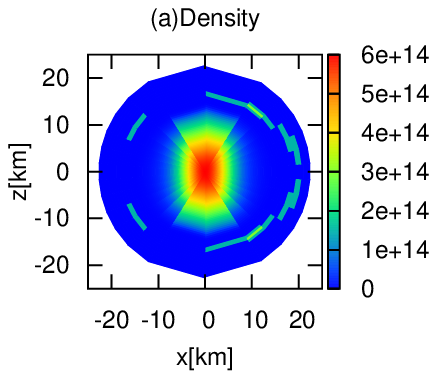}
      \end{minipage}
      \hspace{-1.0cm}
      \begin{minipage}{0.5\hsize}
      \includegraphics[width=12.5cm]{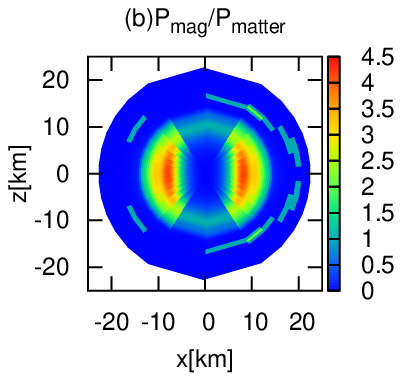}
      \end{minipage}
      \vspace{-0.0cm}
      \\
      \begin{minipage}{0.5\hsize}
      \includegraphics[width=12.5cm]{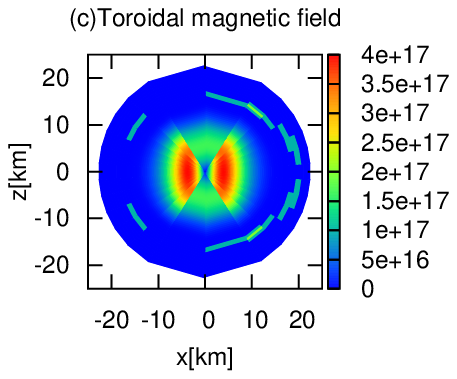}
      \end{minipage}
      \hspace{-0.0cm}
      \begin{minipage}{0.5\hsize}
      \includegraphics[width=10.0cm]{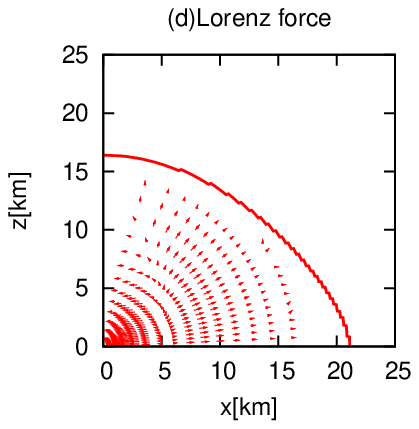}
      \end{minipage}
      \\
      \begin{minipage}{0.5\hsize}
      \includegraphics[width=12.5cm]{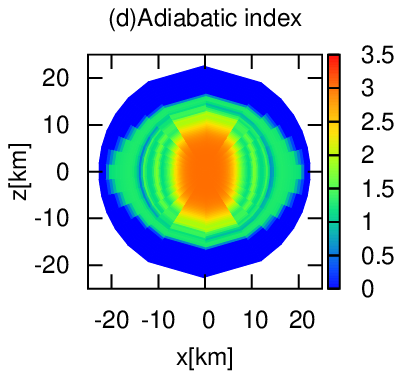}
      \end{minipage}
    \end{tabular}
    \caption{\label{fig:rot}Distribution of the rest mass density, 
    the magnetic field, the magnetic pressure ratio to matter pressure, 
    the Lorenz force, and the adiabatic index for the rotating star with SLy EOS.
    (a) Equi-density contours, (b) equi-$P_{\rm mag}/P_{\rm matter}$ 
    contours, (c) equi-$B_{(\phi)}$ contours, (d) the Lorenz force, and (e) adiabatic index  
    on the meridional plane for the $M=2.00M_\odot$, 
    $B_{\rm max}=5.49\times 10^{17}$G, $R_{\rm cir}=25.0$[km], 
    $\Omega_3=1.87$, and $H/|W|=0.152$ model.
    In the panel (a)-(c), the solid lines 
    correspond to the stellar surface. In the panel of (d), 
    the thick lines represents the surface.
    }
  \end{center}
\end{figure*}

Next, we proceed to the models with rotation. 
In Fig.~\ref{fig:rot}, we display typical distribution of the rotating equilibrium star
 taking a $M=2.00M_\odot$ model with the SLy EOS characterized 
by $\rho_c=6.0\times 10^{14}[{\rm g/cm^3}]$,  $R_{\rm cir}=25.0{\rm km}$, $r_p/r_e=0.68$, $\bar{e}=-0.24$, $T/|W|=4.77\times 10^{-2}$,  $H/|W|=0.152$.
This model has the maximum magnetic field strength of $5.49\times 10^{17}{\rm G}$ 
and rotates in the mass-shedding limit. 
Comparing Fig.~\ref{fig:rot} with Fig.~\ref{fig:non-rot}, we see that 
basic properties of the rotational equilibrium structure are similar to 
those of the static models. Difference between them only appears near 
the equatorial surface of the stars. As shown in Fig.~\ref{fig:rot}, the density 
distribution is stretched from the rotation axis outward due to the centrifugal force 
and the shape of the stellar surface becomes oblate, which can be also seen from  
the value of the axis ratio $r_p/r_e$. On the other hand, 
the equi-density contours deep inside the stars are prolate, as confirmed 
from the value of the mean deformation rate $\bar{e}$. In the central regions, the 
magnetic fastening of the matter generically dominates over the centrifugal forces.  

 It is should be emphasized that we here single out the magnetic field dominated 
models in which  the ratio of the rotational energy to the gravitational energy $T/|W|$ is 
much smaller than the ratio of the magnetic energy to the gravitational energy $H/|W|$ 
even when they rotate at the breakup  angular velocity because we focus on the magnetic 
effects on the stellar structure. In fact, typical values of $H/|W|$ and $T/|W|$ amongst 
the models computed  in this study are orders of $10^{-1}$ and $10^{-2}$, respectively. 
(As we will see later, we cannot pick out such  magnetic field dominated models 
for the supramassive sequences of the magnetized stars with the realistic EOSs.)
The effects of the rotation 
on the stellar structures are therefore secondary in the models shown in this subsection. 
Likewise, the effects of the EOS on the rotational equilibrium sequences
are predominantly determined by the ones in the non-rotating sequences, which we 
mentioned in the previous section.

\subsection{Constant baryon mass sequence}\label{ssec:bcons}
In this subsection, we pay attention to the constant baryon mass sequences 
of the rotating magnetized stars. As discussed in~\citet{Kiuchi:2008ch}, 
it is possible to single out a sequence of the equilibrium stars
 by keeping the baryon rest mass and the magnetic flux constant simultaneously. 
Such equilibrium sequences may model the isolated 
neutron stars that evolve adiabatically losing angular momentum via the 
gravitational/electromagnetic radiations. 
For simplicity, we omit the evolution in the function 
$K$ (Eq. (\ref{eq:I-con1})), which will change as a result of losing angular 
momentum, although there has been no simple way to determine it so far. 
It should be noted that the efficient gravitational radiations 
can be made possible only if the spin and magnetic axes are misaligned
 \citep{Cutler:2002nw}. 
However, including the misalignment to our model is 
extremely difficult because it breaks down the equilibrium condition
due to the gravitational radiation. With the alignment axes, 
 the spin-down via electromagnetic radiations could be possible 
if the outside of the star is non-vacuum \citep{Goldreich:1969}. 
However, the condition that the outside of the star is vacuum is crucial here
for calculating the equilibrium configurations. Furthermore, the 
method to take into account the mixed (poloidal-toroidal) field in general relativistic
equilibrium configuration is not still established as we mentioned 
in Sec.~\ref{sec:Intro}.
 Bearing these caveats, we boldly pay attention to
 the constant baryon mass and magnetic flux sequence obtained from the equilibrium 
configurations to mimic the evolution of the magnetized neutron star in the following. 

As done in~\citet{Cook:1992}, we divide the equilibrium sequences 
of the magnetized stars into two classes, {\it normal} and {\it supramassive} 
equilibrium sequences. In this study, 
the {\it normal} ({\it supramassive}) equilibrium sequence is defined as 
an equilibrium sequence whose baryon rest mass is smaller (larger) than the maximum 
baryon rest mass of the non-magnetized and non-rotating stars. 
In Table~\ref{tab:bmass}, the maximum baryon rest mass of the spherical 
(magnetized rotating) stars, 
which are referred as $M_{0,\text{sph,max}}(M_{0,\text{max}})$, 
with the each EOSs, are given. Note that $M_{0,\text{max}}$ is the 
maximum baryon mass of the magnetized star models calculated in 
this study.

\begin{table}[H]
\centering
\caption{\label{tab:bmass}Maximum baryon mass of the spherical stars and 
the magnetized rotating star with the each EOSs.}
\begin{tabular}{lcccc}
\hline\hline
                                & SLy  & FPS  & Shen & LS   \\
\hline
$M_{0,\text{sph,max}}[M_\odot]$ & 2.43 & 2.11 & 2.81 & 2.23 \\
$M_{0,\text{max}}[M_\odot]$     & 2.82 & 2.46 & 3.36 & 2.62 \\
\hline
\end{tabular}
\end{table}
\subsubsection{Normal sequence}

First, let us consider the normal equilibrium sequences of the magnetized 
rotating stars, choosing $M_0=1.90M_\odot$ stars with the magnetic 
flux of $\Phi_{30}=1$. Here Table~\ref{tab:bmass} tells us 
that all the stars with $M_0=1.90M_\odot$ belong to the normal sequence 
 irrespective of the EOS. The global physical quantities 
for the models with the different EOSs are summarized in Table~\ref{tab:Mb-const-nom-rot} for 
convenience. It should be emphasized that we select the magnetic field-dominant 
sequence $(H/|W|>T/|W|)$ because we are interested in the effect of strong magnetic 
field on the stellar structure (see Table~\ref{tab:Mb-const-nom-rot}).

\begin{figure*}
  \begin{center}
  \vspace*{40pt}
    \begin{tabular}{cc}
      \resizebox{75mm}{!}{\includegraphics{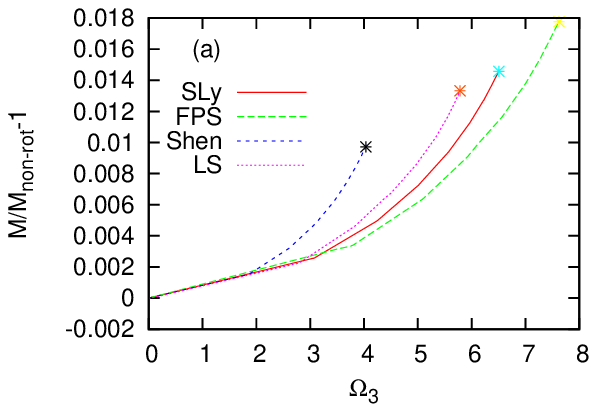}} &
      \resizebox{75mm}{!}{\includegraphics{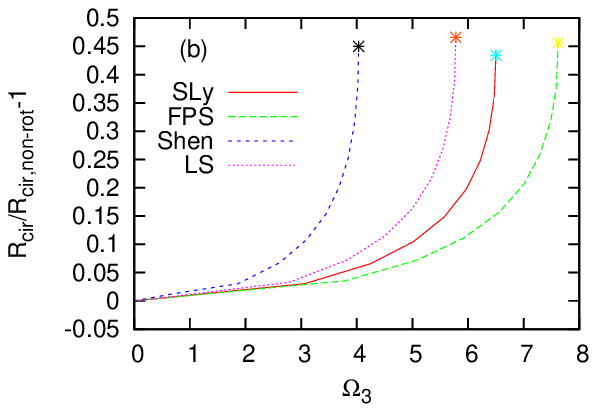}}   \\
      \resizebox{75mm}{!}{\includegraphics{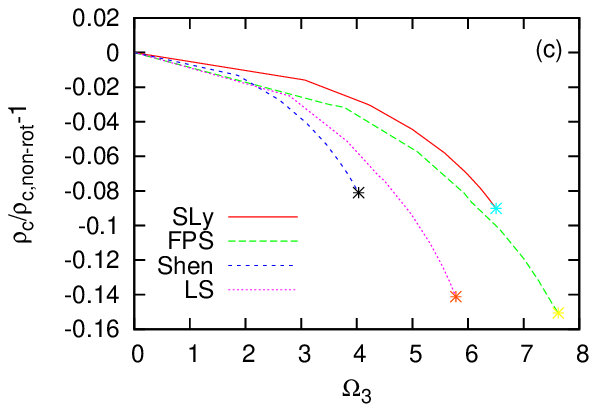}} &
      \resizebox{75mm}{!}{\includegraphics{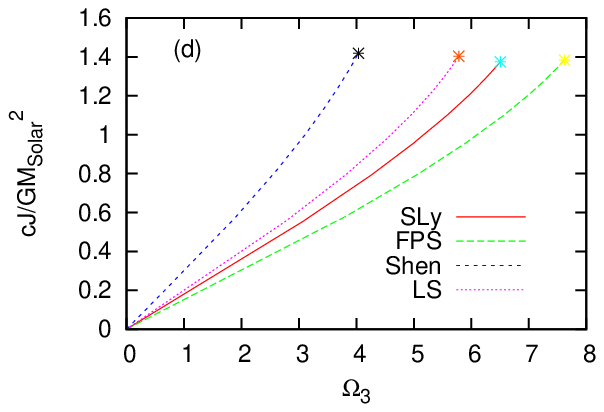}}   \\
      \resizebox{75mm}{!}{\includegraphics{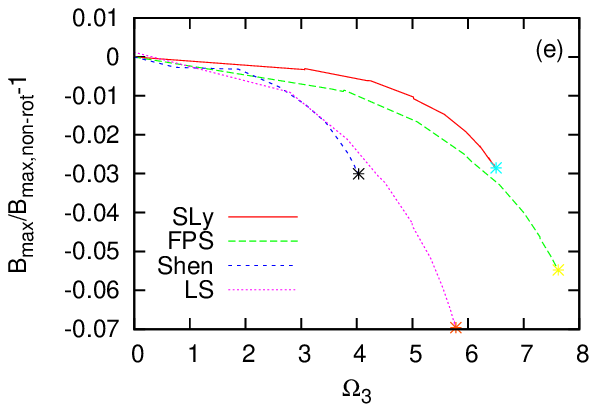}} &
      \resizebox{75mm}{!}{\includegraphics{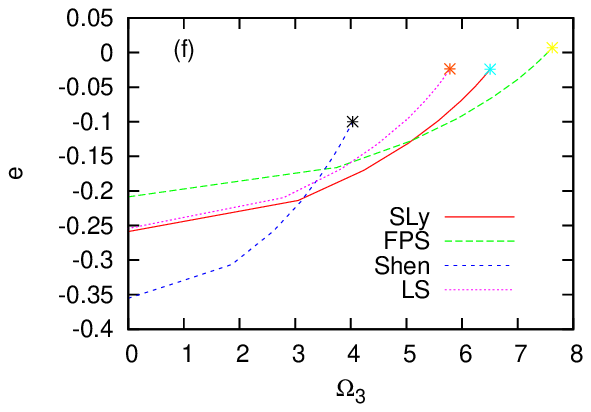}}   \\
    \end{tabular}
    \caption{\label{fig:Mb-Phi-const-nom}
Global physical quantities $M$, $R_{\rm cir}$, $\rho_c$, $J$, $B_{\rm max}$,
and $\bar{e}$ for the constant baryon mass and magnetic flux equilibrium 
sequences with the several realistic EOSs, 
given as functions of $\Omega$. The solid, dashed, long-dashed
, and dotted lines represent the sequences with 
SLy, FPS, Shen, LS EOS, respectively.
All the plots show the increasing (decreasing) rates of the global 
quantities from those for the non-rotating limits. 
All the equilibrium 
sequences are referred to the normal equilibrium sequences characterized
by the constant baryon rest mass $M_0=1.90M_\odot$ and 
the constant magnetic flux $\Phi_{30}=1.0$. The asterisks indicate 
the mass-shedding models. 
    }
  \end{center}
\end{figure*}

In Fig.~\ref{fig:Mb-Phi-const-nom}, the relative change of
 the global physical quantities to those of the spherical stars, such as 
$M$, $R_{\rm cir}$, $\rho_{c,15}$, $J$, $B_{{\rm max},18}$, 
and $\bar{e}$ are given as functions of $\Omega_3$ for the constant baryon mass and 
magnetic flux equilibrium sequences for the different EOSs. Here 
 $\Omega_3$ means the angular velocity normalized by $10^3$[rad/s]. 
It can be seen from panel (a), all the normal equilibrium sequences,
 starting at the non-rotating equilibrium stars ($\Omega_3 = 0$),  
continue to the mass-shedding limits, at which the sequences terminate
(note in each panel, the asterisks indicate the mass-shedding limits).
As the angular velocity increases, 
the gravitational masses and the circumferential radii increase (panels (a) and (b))
 due to the centrifugal forces, 
while the central densities and the maximum amplitudes of the magnetic fields decrease 
(panels (c) and (e)) because of the constancy of the baryon rest mass and the magnetic flux,
 respectively. The angular momenta monotonically increase with the 
angular velocity, which implies that the angular momentum loss via gravitational 
radiation results in the spin down of the star for the normal equilibrium sequences. 
As seen, these qualitative behaviors are found to be common to the realistic EOSs 
employed in this study.

\begin{figure*}
  \begin{center}
  \vspace*{40pt}
    \begin{tabular}{cc}
      \resizebox{75mm}{!}{\includegraphics{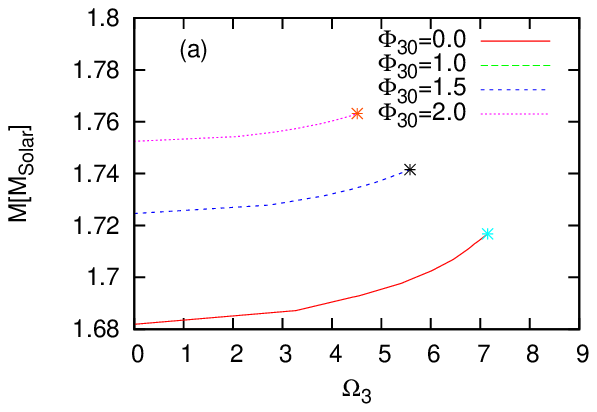}} &
      \resizebox{75mm}{!}{\includegraphics{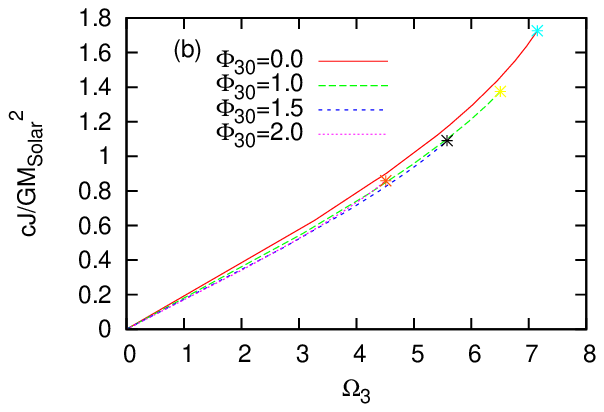}} \\
    \end{tabular}
    \caption{\label{fig:Mb-Phi-const-nom2}
    Global physical quantities $M$ and $J$ for the constant baryon mass and 
    magnetic flux equilibrium sequences with SLy EOS, given as functions of $\Omega$. 
    Results for the $M_0=1.9M_\odot$ models with $\Phi_{30}=0$, $1$, $1.5$, and 
     $2$ are shown. The solid, dashed, long-dashed, and dotted lines correspond to 
     the equilibrium sequences characterized by $\Phi_{30}=0$, $1$, $1.5$, and 
     $2$, respectively. 
    }
  \end{center}
\end{figure*}

Choosing SLy EOS for example, we show in Fig.~\ref{fig:Mb-Phi-const-nom2}
 how the masses and the angular momenta change with the angular velocity along 
with the constant magnetic flux.
In each panel, the non-magnetized case ($\Phi_{30}=0$) is given for the sake of 
comparison.
As seen, qualitative changes 
induced by rotation are basically independent of the value 
of $\Phi_{30}$.  We find that the angular velocity at the mass-shedding point 
becomes small due to the Lorenz force as the magnetic flux increases. 
 Figure~\ref{fig:rot}(d) 
(or Figure~\ref{fig:non-rot}(d)) 
clearly shows that near the stellar surface, the Lorenz force exerted on the matter 
has the same direction with the centrifugal force. Therefore, the 
stars with the strong magnetic 
fields are easy to reach the mass-shedding limit. 

\begin{figure*}
  \begin{center}
  \vspace*{40pt}
    \begin{tabular}{cc}
      \resizebox{75mm}{!}{\includegraphics{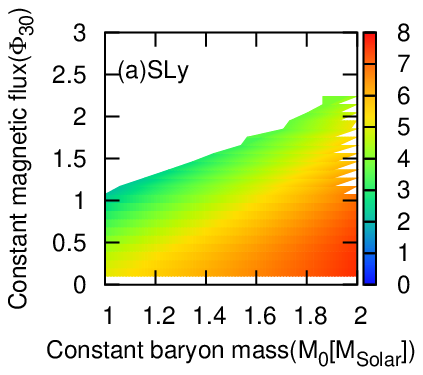}} &
      \resizebox{75mm}{!}{\includegraphics{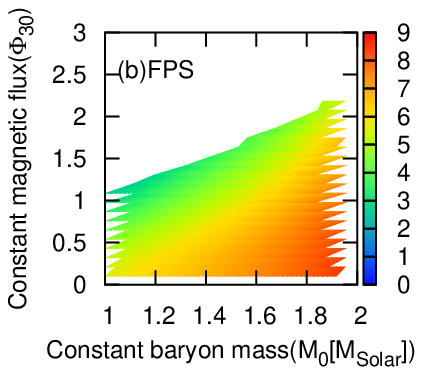}} \\
      \resizebox{75mm}{!}{\includegraphics{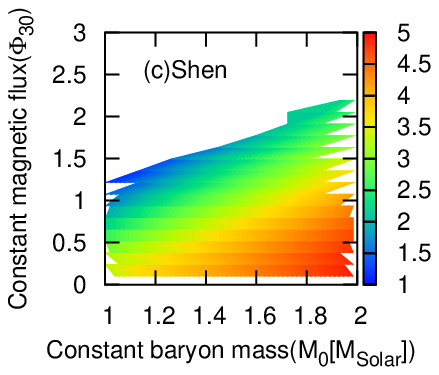}} &
      \resizebox{75mm}{!}{\includegraphics{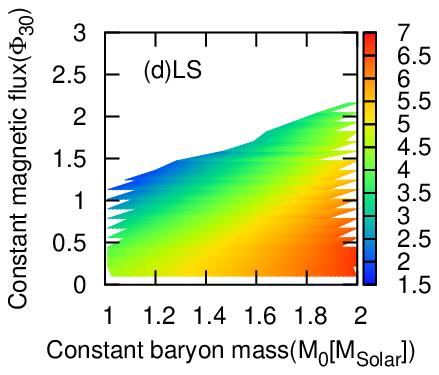}}   \\
    \end{tabular}
    \caption{\label{fig:MS}
     Equi-$\Omega_{\rm ms}$ contours on $M_0$-$\Phi$ plane for (a) SLy, (b) FPS, (c) Shen, 
     and (d) LS EOS. The white regions which lies on the left-upper sides of the panels 
     represent the non-converged limits.
    }
  \end{center}
\end{figure*}

To see clearly the dependence of the mass-shedding angular velocity 
$\Omega_{\rm ms}$ on the baryon mass and magnetic flux characterizing the equilibrium 
sequences, we draw the equi-$\Omega_{\rm ms}$ contours plot on 
$M_0$-$\Phi$ plane in Fig.~\ref{fig:MS}. 
It can be seen that irrespective 
of the EOSs, a large magnetic flux decreases substantially the mass-shedding angular velocities. 
Comparing $\Omega_{\rm ms}$ for the normal equilibrium sequences  with the same baryon mass and magnetic flux, 
we also find that  for the Shen EOS models $\Omega_{\rm ms}$ takes the smallest values, 
for the LS and SLy EOS models middle values, and for the FPS EOS models the largest values 
(see also Fig.~\ref{fig:Mb-Phi-const-nom}(a)). 
This implies we may constrain the EOSs via angular velocity observations. For example, 
if we observe the magnetized neutron star with $M_0=2.0 M_\odot$ and 
$\Omega=8\times10^3{\rm rad/s}$, Shen EOS is rejected because the stars with Shen EOS 
cannot rotate at $\Omega=8\times10^3{\rm rad/s}$ irrespective of the strength of the interior 
magnetic fields. 

\begin{table*}
\centering
\caption{\label{tab:Mb-const-nom-rot}
Global physical quantities for the normal equilibrium sequences of 
the rotating stars with $M_0=1.90M_\odot$ and $\Phi_{30}=1$. 
}
\begin{tabular}{ccccccccc}
\hline\hline
$\rho_{c,15}$&
$M[M_{\rm \odot}]$               &
$R_{\text{cir}}[{\rm km}]$       &
$\Omega_3$          &
$cJ/G M_\odot^2$                 &
$T/|W|$                          &
$H/|W|$                          &
$\bar{e}$                        &
$B_{\text{max},18}$    \\
\hline\hline
\multicolumn{9}{c}{SLy}\\
\hline
 1.21E+00& 1.70E+00& 1.21E+01& 0.00E+00& 0.00E+00& 0.00E+00& 9.40E-02&-2.59E-01&  7.33E-01\\
 1.17E+00& 1.71E+00& 1.28E+01& 4.25E+00& 7.91E-01& 2.27E-02& 9.80E-02&-1.70E-01&  7.29E-01\\
 1.13E+00& 1.72E+00& 1.44E+01& 5.97E+00& 1.21E+00& 5.01E-02& 1.03E-01&-7.09E-02&  7.19E-01\\
 1.11E+00& 1.72E+00& 1.57E+01& 6.38E+00& 1.33E+00& 6.00E-02& 1.04E-01&-3.61E-02&  7.14E-01\\
 1.10E+00& 1.73E+00& 1.73E+01& 6.50E+00& 1.37E+00& 6.33E-02& 1.05E-01&-2.40E-02&  7.12E-01\\
 MS      & -       & -       & -       & -       & -       & -       & -       & -       \\
\hline
\multicolumn{9}{c}{FPS}\\
\hline
 1.75E+00& 1.68E+00& 1.07E+01& 0.00E+00& 0.00E+00& 0.00E+00& 8.51E-02&-2.09E-01& 8.82E-01\\
 1.61E+00& 1.70E+00& 1.18E+01& 5.88E+00& 9.54E-01& 3.38E-02& 9.27E-02&-9.65E-02& 8.61E-01\\
 1.57E+00& 1.70E+00& 1.23E+01& 6.53E+00& 1.09E+00& 4.36E-02& 9.48E-02&-6.55E-02& 8.54E-01\\
 1.52E+00& 1.71E+00& 1.35E+01& 7.30E+00& 1.29E+00& 5.85E-02& 9.80E-02&-1.89E-02& 8.41E-01\\
 1.49E+00& 1.71E+00& 1.56E+01& 7.62E+00& 1.38E+00& 6.67E-02& 9.97E-02& 7.00E-03& 8.34E-01\\
 MS      & -       & -       & -       & -       & -       & -       & -       & -       \\
\hline
\multicolumn{9}{c}{Shen}\\
\hline
 5.47E-01& 1.75E+00& 1.66E+01& 0.00E+00& 0.00E+00& 0.00E+00& 1.04E-01&-3.55E-01&  4.22E-01\\
 5.32E-01& 1.75E+00& 1.77E+01& 2.61E+00& 8.18E-01& 2.03E-02& 1.07E-01&-2.58E-01&  4.19E-01\\
 5.18E-01& 1.76E+00& 1.91E+01& 3.44E+00& 1.14E+00& 3.76E-02& 1.10E-01&-1.80E-01&  4.15E-01\\
 5.08E-01& 1.76E+00& 2.08E+01& 3.85E+00& 1.32E+00& 4.97E-02& 1.11E-01&-1.28E-01&  4.12E-01\\
 5.03E-01& 1.77E+00& 2.29E+01& 4.02E+00& 1.41E+00& 5.59E-02& 1.12E-01&-1.02E-01&  4.10E-01\\
 MS      & -       & -       & -       & -       & -       & -       & -       & -       \\
\hline
\multicolumn{9}{c}{LS}\\
\hline
 1.18E+00& 1.71E+00& 1.28E+01& 0.00E+00& 0.00E+00& 0.00E+00& 9.14E-02&-2.55E-01&  6.65E-01\\
 1.15E+00& 1.72E+00& 1.33E+01& 2.79E+00& 5.61E-01& 1.11E-02& 9.35E-02&-2.10E-01&  6.59E-01\\
 1.07E+00& 1.73E+00& 1.49E+01& 4.99E+00& 1.11E+00& 4.12E-02& 9.92E-02&-9.80E-02&  6.37E-01\\
 1.03E+00& 1.73E+00& 1.63E+01& 5.54E+00& 1.30E+00& 5.47E-02& 1.02E-01&-5.02E-02&  6.26E-01\\
 1.01E+00& 1.74E+00& 1.88E+01& 5.78E+00& 1.40E+00& 6.23E-02& 1.03E-01&-2.37E-02&  6.19E-01\\
 MS      & -       & -       & -       & -       & -       & -       & -       & -       \\
\hline
\end{tabular}
\end{table*}

\subsubsection{Supramassive sequence}
\begin{figure*}
  \begin{center}
  \vspace*{40pt}
    \begin{tabular}{cc}
      \resizebox{75mm}{!}{\includegraphics{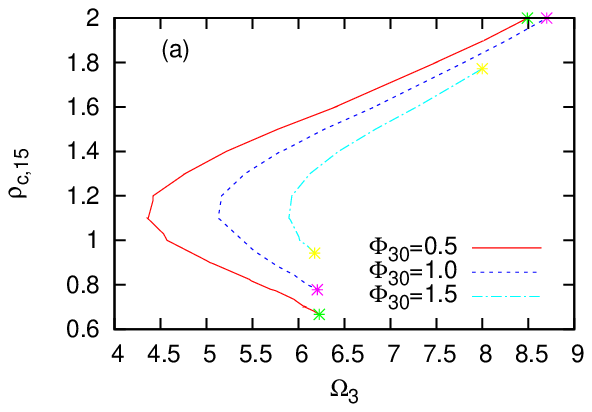}} &
      \resizebox{75mm}{!}{\includegraphics{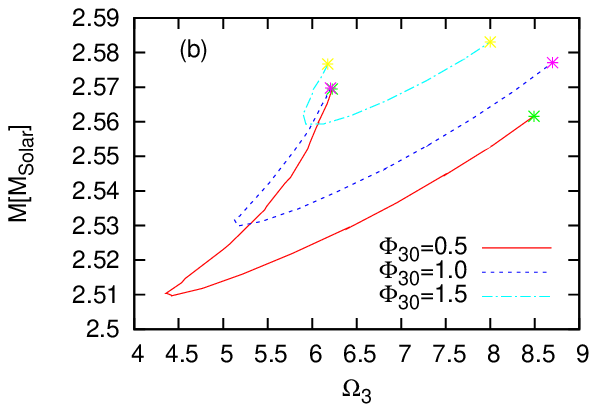}} \\
      \resizebox{75mm}{!}{\includegraphics{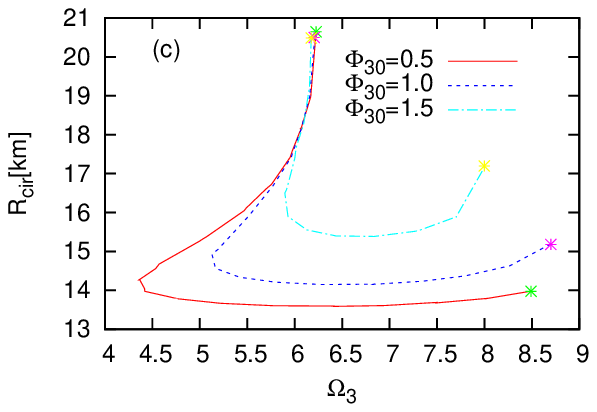}} &
      \resizebox{75mm}{!}{\includegraphics{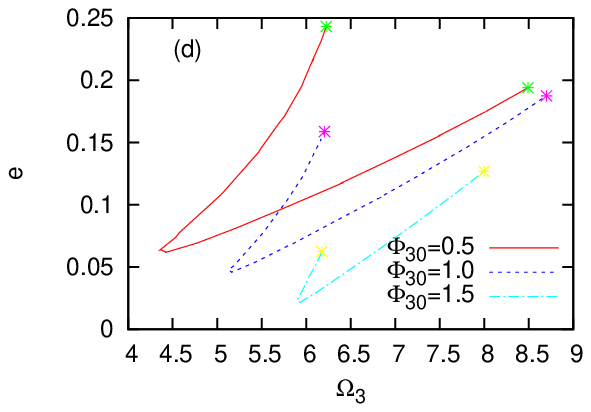}} \\
    \end{tabular}
    \caption{\label{fig:sup-Shen}
Global quantities, $\rho_c$, $\bar{e}$, $M$, and $R_{\rm cir}$ 
along the supramassive sequence 
with Shen EOS  which is characterized
by the constant baryon rest mass $M_0=2.90M_\odot$. The solid, dashed, and long-dashed 
lines correspond to the sequences with $\Phi_{30}=0.5$, $1.0$, $1.5$, respectively.
    }
  \end{center}
\end{figure*}

Now we move on to consider the supramassive equilibrium sequences 
of the magnetized rotating stars. As in the case of the normal sequence, 
the qualitative features in the sequences are not sensitive to the difference
 in the EOSs. So we take the model with Shen EOS in the following.
Figure \ref{fig:sup-Shen} gives the plots 
of the global stellar quantities as a function of $\Omega$ for a constant 
mass (:supramassive $M_0=2.90M_\odot$) and magnetic flux with the Shen EOS. 
Each curve is labeled by its value of $\Phi_{30}$ which is held constant along 
the sequence. 
Values of the global physical quantities, $\rho_{c}$, $M$, $R_{\rm cir}$, 
$\Omega$, $J$, $T/|W|$, $H/|W|$, $\bar{e}$, and $B_{{\rm max},18}$, for some selected 
models are summarized for convenience in Table~\ref{tab:Mb-const-sup-rot}. 

From panel (a), the supramassive sequences begin at the mass-shedding limits with 
lower central density $\rho_{c}$ as the angular velocity $\Omega$ 
decreases and reach the turning points (the slowest rotation points), then move to the other mass-shedding limits with 
higher central density $\rho_{c}$ as the angular velocity increases.
The gravitational mass decreases at first, 
and encounters the turning point, then it begins to increase 
as the central densities decrease  
(panel (b)). The circumferential radius decreases because of the increase of the central 
density and keeps the almost constant value after the turning point (panel (c)). 
From Figs.~\ref{fig:sup-Shen}(d), we find most models of the supramassive 
sequences have the positive values of the mean deformation rate, which implies 
the stars oblately deform. This feature can be also seen in the sequences with 
the other EOSs. 

\begin{figure*}
  \begin{center}
  \vspace*{40pt}
    \begin{tabular}{cc}
      \resizebox{75mm}{!}{\includegraphics{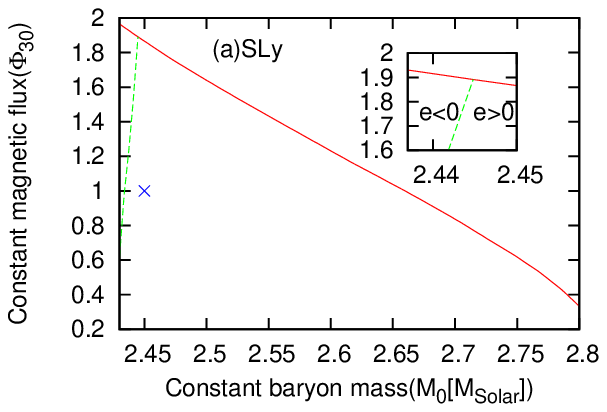}} &
      \resizebox{75mm}{!}{\includegraphics{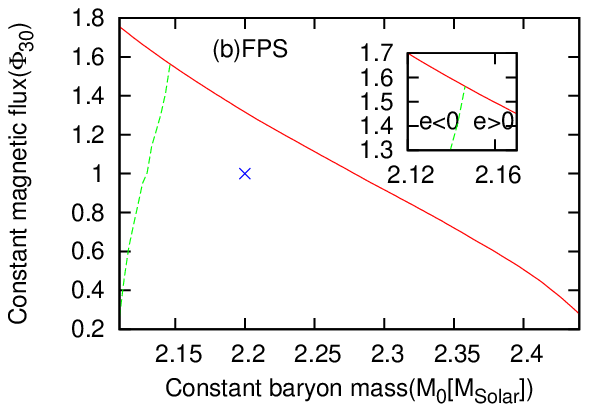}} \\
      \resizebox{75mm}{!}{\includegraphics{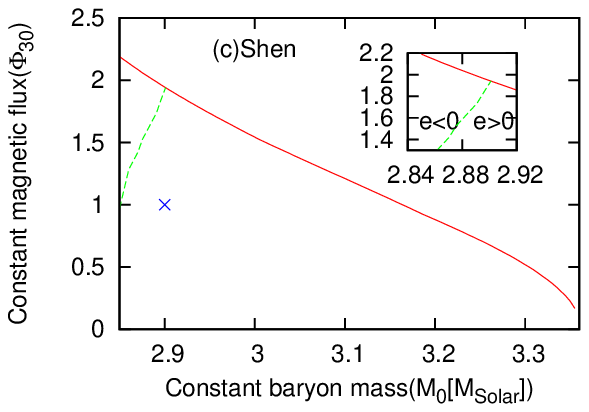}} &
      \resizebox{75mm}{!}{\includegraphics{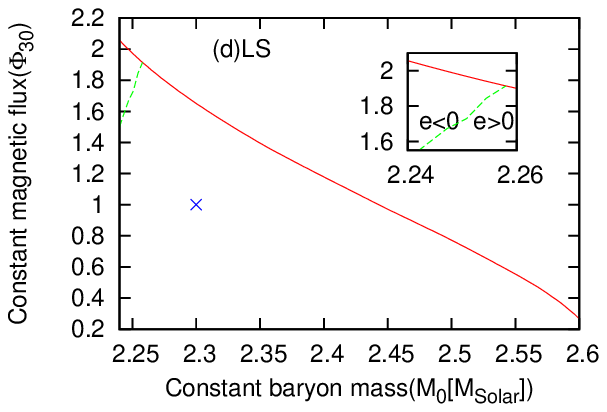}}   \\
    \end{tabular}
    \caption{\label{fig:prol}
    Phase diagram on constant baryon mass and magnetic flux plane of 
    the supramassive sequences for (a)SLy, 
    (b) FPS, (c) Shen, and (d) LS EOS. The solid lines correspond to the 
    critical line, above which there exists no physical solution (see 
    the text for details).    The dashed lines show the models with $\bar{e}=0$. 
    In the left (right) side regions of the dotted lines, 
    the supramassive sequences have the prolate (oblate) solutions. 
    The magnified panels are put on the upper right of the each figures.
    The cross symbols in each panel represent the sequences given in 
    Table~\ref{tab:Mb-const-sup-rot} (see the text for details).  }
  \end{center}
\end{figure*}

In fact, Table~\ref{tab:Mb-const-sup-rot} shows that 
the rotational energy dominates over the magnetic energy $(T/|W|>H/|W|)$ in 
these sequences. This means that we pick out the rotation-dominated 
sequence \citep{Yoshida:2006a,Yoshida:2006b} for Table 4 and Figure 11. 
However, it should be noted that it depends on the values of the baryon rest mass and 
magnetic flux whether a sequence belongs to the rotation-dominated one or 
the magnetic-field-dominated one. 
To get an overall picture, we show the phase diagram in Figure~\ref{fig:prol} 
on $M_0$ and $\Phi_{30}$ plane of the supramassive sequence with each realistic 
EOS. The solid red line in Fig.~\ref{fig:prol} indicates 
the boundary, above which
 there is no physical solution (see, e.g.,\citet{Kiuchi:2008ch}).
The regions below the solid lines correspond to a supramassive sequence. 
We plot the sequences given in 
Table~\ref{tab:Mb-const-sup-rot} as the cross symbols in each panel. 
In the right side regions of the dotted lines in Fig.~\ref{fig:prol}, 
the sequences only have the oblate $(\bar{e}>0)$ stars. On the other hand, 
in the left side regions, there exit the sequences which have the 
prolate $(\bar{e}<0)$ stars (see also the magnified panels 
in the each plots). Figure~\ref{fig:prol} clearly shows that 
the supramassive sequence with the realistic EOSs almost belong to 
the rotation-dominated sequence because the magnetic-field-dominated 
sequences are easy to be prolately deformed as discussed in Sec.~\ref{ssec:rot}. 
 Looking carefully, it can be seen that 
for FPS EOS, the parameter region which permits 
the prolately deformed stars is widest. For Lattimer-Swesty EOS, the region 
is narrowest vice versa and for SLy and Shen EOS, it is in medium.
It should be mentioned that 
for the polytropic EOS, the magnetic-field-dominated sequences along the supramassive 
sequences easily appear (\citet{Kiuchi:2008ch}) and this difference may be useful 
for the constraining EOS via gravitational wave observations as we will discuss 
in Sec.~\ref{sec:summary}. 

\begin{figure*}
  \begin{center}
  \vspace*{40pt}
    \begin{tabular}{cc}
      \resizebox{75mm}{!}{\includegraphics{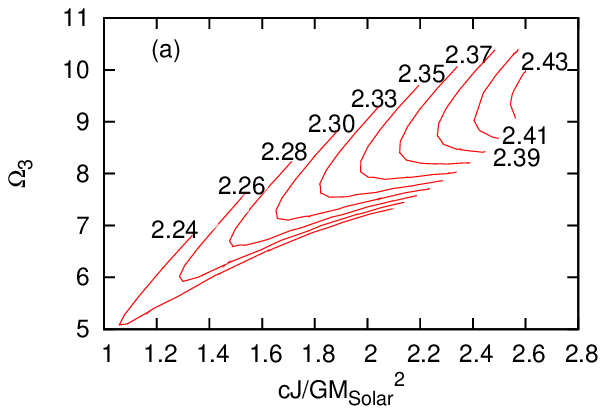}} &
      \resizebox{75mm}{!}{\includegraphics{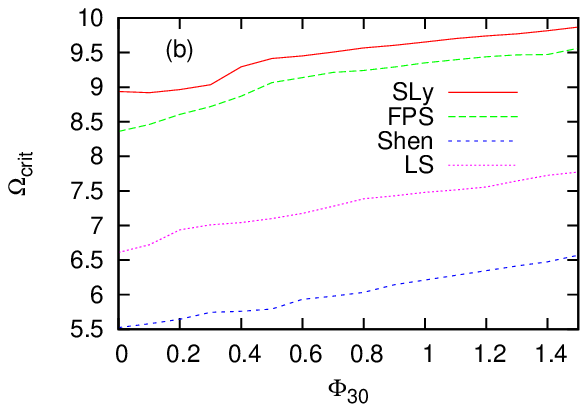}} \\
    \end{tabular}
    \caption{\label{fig:spinup}
    (a) Angular momentum and angular velocity relation in 
    the supramassive sequences with LS EOS and $\Phi_{30}=1.0$. 
    The numbers attached on the each sequences represent the 
    values of constant baryon rest mass in units of $M_\odot$.
    (b) Critical spin up angular velocity as functions of the 
    magnetic flux for each EOS. The critical angular velocity 
    is defined by the value from which the stars start to spin up 
    (see the text in details). 
    }
  \end{center}
\end{figure*}

Finally, let us investigate the spin up of the stars 
as the stellar angular momentum decreases, which is 
one of the astrophysically important features of the constant baryon mass
equilibrium sequences of the non-magnetized rotating stars (see, e.g. 
\citet{Shapiro:1990} and \citet{Cook:1992}). 
This spin up effect of the relativistic stars containing purely poloidal
magnetic fields and purely toroidal magnetic fields has also been found by~\citet{Bocquet:1995} 
and~\citet{Kiuchi:2008ch}, respectively. In this study, 
we find that the the spin up effects occur for the stars containing purely toroidal magnetic fields 
with the realistic equations  of state. Figure~\ref{fig:spinup}(a) shows the 
angular velocity $\Omega$ as a function of the angular momentum $J$ 
along the supramassive sequences with LS EOS. These sequences are characterized 
by the constant magnetic flux $\Phi_{30}=1.0$ and the number attached by  
each curve indicates the constant baryon mass values in units of 
$M_\odot$. 
We observe the spin up effect, the angular velocity increase as the 
angular momentum decreases, in the supramassive sequences with $2.39$, $2.41$, 
and $2.43M_\odot$.
This spin up is induced by the decrease of the moment of inertia 
with decreasing the angular momentum. 
Reaching the turning point of the constant sequence path of Fig. 13, 
 the stars in absent of the magnetic fields were pointed out to begin to collapse 
to black hole (\citet{Cook:1992} according to the criterion by \citet{Friedman:1988}). 
Qualitatively the magnetized stars here are also expected to form black holes 
on the analogy, but to find the criterion for the magnetized case needs further 
investigation. 
We here define a critical angular velocities, $\Omega_{\rm up}$,  
with which the stars can start to spin up.
 Finally, we get the relationships between $\Omega_{\rm up}$ and $\Phi$ 
for the different EOSs, which are drawn in Fig.~\ref{fig:spinup}(b). 
Our analysis indicates that the hierarchy of this spin-up angular velocity is 
$\Omega_{\rm up,SLy} > \Omega_{\rm up,FPS} > \Omega_{\rm up,LS}>\Omega_{\rm up,Shen}$ 
and this relation holds even if the sequences have strong magnetic fields. 
Furthermore the values of $\Omega_{\rm up}$ are found to
 increase with the constant magnetic flux 
irrespective of the EOSs and the lines for the different EOSs never cross each other.

\begin{table*}
\centering
\begin{minipage}{140mm}
\caption{\label{tab:Mb-const-sup-rot}
Global physical quantities for the supramassive equilibrium sequences 
of the rotating stars.
}
\begin{tabular}{ccccccccc}
\hline\hline
$\rho_{c,15}$ &
$M[M_{\rm \odot}]$                &
$R_{\text{cir}}[{\rm km}]$        &
$\Omega_{3}$         &
$cJ/G M_\odot^2$                  &
$T/|W|$                           &
$H/|W|$                           &
$\bar{e}$                         &
$B_{\text{max},18}$     \\
\hline\hline
\multicolumn{9}{c}{SLy $(M_0,\Phi_{30})=(2.45M_\odot,1.0)$}\\
\hline
 MS      & -       & -       & -       & -       & -       & -       & -       & -        \\
 1.27E+00& 2.14E+00& 1.51E+01& 8.95E+00& 2.89E+00& 1.10E-01& 2.02E-02& 2.26E-01&  3.93E-01\\
 1.50E+00& 2.11E+00& 1.18E+01& 7.69E+00& 2.08E+00& 6.14E-02& 1.63E-02& 1.26E-01&  4.01E-01\\
 2.00E+00& 2.08E+00& 1.02E+01& 4.67E+00& 1.04E+00& 1.65E-02& 1.15E-02& 2.39E-02&  4.11E-01\\
 2.60E+00& 2.11E+00& 9.79E+00& 9.29E+00& 1.96E+00& 4.95E-02& 9.28E-03& 1.04E-01&  4.30E-01\\
 2.80E+00& 2.13E+00& 9.85E+00& 1.11E+01& 2.35E+00& 6.83E-02& 8.68E-03& 1.46E-01&  4.35E-01\\
 3.00E+00& 2.15E+00& 1.00E+01& 1.27E+01& 2.70E+00& 8.64E-02& 8.10E-03& 1.86E-01&  4.39E-01\\
 MS      & -       & -       & -       & -       & -       & -       & -       & -        \\
\hline
\multicolumn{9}{c}{FPS $(M_0,\Phi_{30})=(2.20M_\odot,1.0)$}\\
\hline
 MS      & -       & -       & -       & -       & -       & -       & -       & -        \\
 1.58E+00& 1.93E+00& 1.40E+01& 9.52E+00& 2.29E+00& 1.04E-01& 2.38E-02& 2.13E-01&  4.60E-01\\
 1.80E+00& 1.92E+00& 1.15E+01& 8.97E+00& 1.91E+00& 7.53E-02& 2.08E-02& 1.55E-01&  4.74E-01\\
 2.30E+00& 1.90E+00& 1.01E+01& 8.20E+00& 1.48E+00& 4.65E-02& 1.62E-02& 9.31E-02&  4.99E-01\\
 2.60E+00& 1.90E+00& 9.74E+00& 8.84E+00& 1.52E+00& 4.81E-02& 1.46E-02& 9.90E-02&  5.13E-01\\
 2.80E+00& 1.90E+00& 9.62E+00& 9.69E+00& 1.64E+00& 5.44E-02& 1.37E-02& 1.15E-01&  5.22E-01\\
 3.00E+00& 1.91E+00& 9.57E+00& 1.08E+01& 1.79E+00& 6.34E-02& 1.30E-02& 1.36E-01&  5.31E-01\\
 MS      & -       & -       & -       & -       & -       & -       & -       & -        \\
\hline
\multicolumn{9}{c}{Shen $(M_0,\Phi_{30})=(2.9M_\odot,1.0)$}\\
\hline
 MS      & -       & -       & -       & -       & -       & -       & -       & -        \\
 7.78E-01& 2.57E+00& 2.05E+01& 6.21E+00& 3.80E+00& 9.32E-02& 5.16E-02& 1.59E-01&  4.38E-01\\
 8.04E-01& 2.56E+00& 1.82E+01& 6.08E+00& 3.60E+00& 8.47E-02& 5.00E-02& 1.40E-01&  4.43E-01\\
 1.19E+00& 2.53E+00& 1.46E+01& 5.16E+00& 2.40E+00& 4.02E-02& 3.78E-02& 4.59E-02&  4.92E-01\\
 1.40E+00& 2.53E+00& 1.42E+01& 5.80E+00& 2.58E+00& 4.51E-02& 3.47E-02& 6.72E-02&  5.14E-01\\
 1.80E+00& 2.56E+00& 1.44E+01& 7.80E+00& 3.38E+00& 7.18E-02& 3.11E-02& 1.46E-01&  5.53E-01\\
 2.00E+00& 2.58E+00& 1.52E+01& 8.70E+00& 3.78E+00& 8.65E-02& 2.99E-02& 1.88E-01&  5.69E-01\\
 MS      & -       & -       & -       & -       & -       & -       & -       & -        \\
\hline
\multicolumn{9}{c}{LS $(M_0,\Phi_{30})=(2.3M_\odot,1.0)$}\\
\hline
 MS      & -       & -       & -       & -       & -       & -       & -       & -        \\
 1.14E+00& 2.05E+00& 1.66E+01& 7.65E+00& 2.58E+00& 1.02E-01& 2.29E-02& 2.19E-01&  3.52E-01\\
 1.30E+00& 2.03E+00& 1.35E+01& 7.24E+00& 2.13E+00& 7.38E-02& 2.03E-02& 1.59E-01&  3.70E-01\\
 1.80E+00& 2.01E+00& 1.13E+01& 5.95E+00& 1.38E+00& 3.27E-02& 1.52E-02& 6.42E-02&  4.10E-01\\
 2.00E+00& 2.00E+00& 1.10E+01& 6.14E+00& 1.35E+00& 3.11E-02& 1.40E-02& 6.22E-02&  4.24E-01\\
 2.30E+00& 2.01E+00& 1.07E+01& 7.32E+00& 1.54E+00& 3.92E-02& 1.26E-02& 8.55E-02&  4.44E-01\\
 2.40E+00& 2.01E+00& 1.06E+01& 7.91E+00& 1.65E+00& 4.41E-02& 1.22E-02& 9.85E-02&  4.50E-01\\
 MS      & -       & -       & -       & -       & -       & -       & -       & -        \\
\hline
\end{tabular}
\end{minipage}
\end{table*}

\section{Discussion and Summary}\label{sec:summary}

\subsection{Discussion}
We give discussions in this subsection, paying attention to 
the properties of the strongly magnetized neutrons stars with the realistic EOSs.

(1) As mentioned, the magnetized neutron stars could evolve adiabatically 
losing its 
angular momentum by gravitational or electromagnetic radiation. 
Equations (2.9) and (2.10) in~\citet{Cutler:2002nw} give us the order 
estimation of angular momentum loss time scale by gravitational radiation $\tau_{\rm GW}$ 
and that by electromagnetic radiation $\tau_{\rm EM}$ as 
$\tau_{\rm GW}/\tau_{\rm EM} = 8.9\times10^{-4}(B_{\rm d}/10^9 {\rm G})^2(\epsilon_B/10^{-7})^{-2}
(\nu_s/{\rm kHz})^{-2}$, where $\epsilon_B$, $B_d$, and $\nu_s$ represent 
the degree of deformation induced by the toroidal magnetic field, the dipole magnetic field outside 
the star, and the rotation frequency. If we assume $B_{\rm d}=10^{14}{\rm G}$ and $\epsilon_B=10^{-7}$, 
the equation implies the angular momentum loss is driven by the electromagnetic radiation. 
On the other hand, our models suggest that $\epsilon_B$ can reach to the order of 0.1 as we discussed in 
Sec.~4. In this case, the gravitational radiation could be main agent of the angular momentum loss. 

As mentioned, the mean deformation rates $\bar{e}$ for the strongly
magnetized stars depend on the parameters of the equilibrium models. 
Along the normal sequences, they are basically negative, 
which means that the mean matter distributions are prolate, even when the stars
rotate at nearly the breakup angular velocity (see panel (f) of Fig.~\ref{fig:Mb-Phi-const-nom}). 
High toroidal fields enough to make the equilibrium configuration prolate in the absence 
of rotation, are pointed out 
 to be a good emitter of the gravitational waves \citep{Cutler:2002nw}, 
leading to a secular instability, in which the wobbling 
angles between the rotation axis and the star's magnetic axis would grow on the
 dissipation timescale, until they become orthogonal. 

Following to Eq.~(4.2) in~\citet{Cutler:2002nw}, 
${S}/{N} = 11.7 \left({10{\rm kpc}}/{D}\right)\left({\epsilon_B}/{10^{-6}}\right)
\left({10^{14}{\rm G}}/{B_d}\right)\left[\ln\left({f_{\rm max}}/{f_{\rm min}}\right)\right]$, 
we may estimate the signal 
to noise ratio of gravitational wave from the magnetized neutron star, 
where $D$ and $f_{\rm max}(f_{\rm min})$ represent the distance to the source 
and the maximum (minimum) rotation frequency. 
 Here, it is assumed that the neutron star spin-downs due to the magnetic dipole 
radiation, though in the present neutron star models, we fully omit the exterior 
dipole magnetic field $B_d$ assuming it does not affect the neutron star structures.
In our models, we find the maximum value of $\epsilon_B$ is of an order of $10^{-1}$, 
if the magnetic fields nearly reach the equipartition. 
Supposing the dipole magnetic field is $10^{14}$G, we have $S/N\sim 10$ for $D=1\ {\rm Gpc}$. 
To constrain the EOSs by the gravitational wave observations, we need to investigate 
the gravitational waveforms in details using the presented models here, which will be 
presented soon elsewhere.


(2) The stars in the supramassive sequences with the sufficient large baryon mass 
start to spin up losing their angular momentum, irrespective of the EOSs. The critical angular velocities 
at which the spin-up begins depend on the specified constant magnetic flux 
and the selected EOS. Let us discuss the possibility of constraining the equation 
of state by the spin-up effect. If we detect the spin-up effect in a magnetized 
neutron star and estimate its critical angular velocity as e.g., $8\times10^3$[rad/s], 
SLy and FPS EOS would be rejected because the spin-up never occurs in 
the stars with these two EOSs. 
 In our models which can spin up losing their angular momentum, the ratio of the rotational energy to 
the gravitational biding energy $T/|W|$ never excess $0.27$, which is the 
conservative value of the dynamical instability. This result implies 
the strongly magnetized stars may survive during its spinning up. 

(3) The gravitational masses of the strongly magnetized stars constructed in this 
study are seemingly too large (see Table~\ref{tab:Mb-const-sup-rot}) because the 
canonical value of the neutron star mass is $1.4M_\odot$. However, 
the population synthesis study in ~\citet{Ferrario:2007bt} 
indicates the neutron stars can possess the heavy mass $\sim 2 M_\odot$ if 
they have the strong magnetic fields. Thus, our strongly magnetized neutron stars 
could be models of the magnetar and the high field neutron star. 
In combination with (1), we may obtain the 
information about the field strength from the observation of the gravitational waves.

(4) We shall comment on the definition of the supramassive and 
the normal sequences for the magnetized stars, which are categorized by the 
baryon mass following \citet{Cook:1992} in this study. 
For the non-magnetized stars, a characteristic feature of the normal sequences is that 
they always begin at a non-rotating solution and end at a mass-shedding solution. 
For the magnetized stars, on the other hand, we find that some normal sequences 
do not start at a non-rotating solution and include no non-rotating solutions. This 
implies that for the magnetized stars, 
we have yet an another option to define the supramassive and the normal sequences. 
The definition can be given by the way how the equilibrium sequences end. 
If we define the supramassive sequences as the sequence that terminates 
at the mass-shedding solutions, we could extend the supramassive 
sequences to the lower mass regions (see Fig.~\ref{fig:prol}).  
As can be expected from Fig.~\ref{fig:prol}, these extended supramassive sequences 
allow more prolate solutions, which should be inevitably magnetic-field dominated. 
However seeking the solutions in such parameter regime is hindered due to the
 non-convergence limits. We are now undertaking to update our numerical 
scheme to make the phase diagram of Fig.~\ref{fig:prol} more comprehensive.

(5)Recently, \citet{Kiuchi:2008ss} have performed axisymmetric stability analysis of 
the toroidal magnetic field by making the GRMHD simulation 
and found the magnetic field with $k\ne1$ is unstable. 
Therefore, we limit ourselves to $k=1$ case in this work. 
As for non-axisymmetric stability, 
Tayler has shown the toroidal magnetic field with $k=1$ 
induces the kink instability near the magnetic axis in the Newtonian 
analysis~\citep{Tayler: 1973}. 
Therefore, some models obtained in this study might be unstable due to the kink instability. 
However, the perturbation analysis only predicts the occurrence of instability. 
It is still unclear how the magnetic field evolves after the 
Tayler instability sets in. Will they decay or settle down 
to ``new'' equilibrium state like the axisymmetric case \citep{Kiuchi:2008ss}? 
To draw a robust conclusion, the 3D GR MHD simulations are necessary, which 
is our future work. 

(6) As mentioned in section 1.1, the strong 
magnetic fields inside neutron stars should decay during their 
evolutions, which has been totally neglected in this study.
Employing Eq. (61) in \citet{Goldreich},  the lifetime 
(decaying timescales of the magnetic fields) for the presented models here
can be estimated as an order of $\sim 10^{4}$ years. 
Although more rapid decay is thought 
to be possible by taking into account the neutron star crusts \citep{pons}, 
 the lifetime seems not so contradictory with  
 the observations implying the lifetimes less than $10^{5}$ years \citep{wod}.

(7) In this paper, we considered only the magnetic effects on the
equilibrium configurations through the Lorentz force but did not take 
into account the additional changes caused by the magnetic effects on
the EOS. For the super-strong magnetic fields of $B > 10^{5} B_{\rm
QED}$, there appear two counter effects in the EOSs, namely the
stiffening and softening due to the anomalous magnetic
moments of the nucleons and to the Landau quantization, respectively
\citep{brod}. Since the equilibrium configurations
are determined by the subtle local balance of the pressure gradient, 
vs. predominantly the gravitational force (plus the Lorentz force and
the centrifugal force), it is by no means a trivial
problem to see the effects of the local change of the pressure 
on the important global properties of the strongly magnetized
equilibrium star, such as the mass and the radius. 
To answer this important problem, it is indispensable to
 incorporate the magnetic corrections to the EOSs, however beyond 
scope of this paper.

\subsection{Summary}
In this study, we have investigated equilibrium sequences of relativistic stars containing purely toroidal magnetic fields with four kinds of realistic EOSs of SLy, FPS, Shen, 
and LS, which have been often employed in recent MHD studies relevant for 
magnetized neutron stars.  Solving master equations numerically using the 
Kiuchi-Yoshida scheme, we have constructed thousands of equilibrium configurations 
in order to study the effects of the realistic EOSs. 
 Particularly we have paid attention to the equilibrium sequences of constant 
baryon mass and/or constant magnetic flux, which model evolutions of an isolated neutron
star, losing angular momentum via gravitational waves. 
Important properties obtained in this study are summarized as follows ; 
(1) Along the maximum gravitational mass sequences, the stars with the realistic EOSs cannot be prolate as much as the stars with a polytropic EOS of $n=1$ with $n$ being the 
polytropic index, which has been often employed to mimic the nuclear EOSs. 
(2) The dependence of the mass-shedding angular velocity on the EOSs 
 along a constant baryon mass and magnetic flux sequences, 
is determined from that of the non-magnetized case. Along the sequence, the stars with 
Shen(FPS) EOS reach the mass-shedding limit 
at the smallest(largest) angular velocity, while the stars with 
SLy or Lattimer-Swesty EOSs take the moderate values.
(3)  For the supramassive sequences, the equilibrium configurations 
become generally oblate for the realistic EOSs, although 
the prolately deformed stars can exist in a narrow parameter region spanned by the
 constant baryon mass and magnetic flux. 
For FPS EOS, the parameter region which permits 
the prolately deformed stars is widest. For Lattimer-Swesty EOS, the region 
is narrowest vice versa and for SLy and Shen EOS, it is in medium. 
(4) For the supramassive sequences, the angular velocities $\Omega_{\rm up}$,  
above which the stars start to spin up as they lose angular momentum, are found to 
depend sharply on the realistic EOSs. 
The hierarchy of the critical spin up angular velocity is 
SLy $>$ FPS $>$ LS $>$ Shen EOS and this turn never change even if they have strong magnetic fields. Our results suggest the relativistic stars containing purely toroidal 
magnetic fields will be a potential source of gravitational waves 
and the EOSs within such stars can be constrained by observing the angular velocities,
 the gravitational waves, and the signature of the spin up. 

\acknowledgments
K. Kiuchi expresses thanks to K. Maeda and S. Yamada for invaluable discussion. 
K.Kotake expresses thanks to K. Sato for continuing encouragements.
Numerical computations were in part carried on XT4 and 
general common use computer system at the center for Computational Astrophysics, CfCA, the National Astronomical Observatory of Japan.  This
study was supported in part by the Grants-in-Aid for the Scientific Research 
from the Ministry of Education, Science and Culture of Japan (Nos. 19540309 and 20740150).

\begin{thebibliography}{}

\bibitem[Akiyama et al.(2003)]{Akiyama:2002xn}
  Akiyama,~S., Wheeler,~J.~C., Meier,~D.~L., \& Lichtenstadt,~I. 
  2003, APJ, 584, 954

\bibitem[Akmal \& Pahdharipahde(1998)]{akmal}
  Akmal, A., Pandharipande, V.~R., \& Ravenhall, D.~G. 
  1998, PRC, 58, 1804 

\bibitem[Balbus \& Hawley(1991)]{Balbus:1991}
  Balbus,~S.~A. \& Hawley,~J.~F. 
  1991, APJ, 376, 214

\bibitem[Baym \& Pethick(1979)]{bp}
  Baym, G., \& Pethick, C. 
  1979, Ann. Rev. Astron. Astrophys., 17, 415 

\bibitem[Boucquet et al.(1995)]{Bocquet:1995} 
  Bocquet,~M., Bonazzola,~S., Gourgoulhon,~E., \& Novak,~J. 
  1995, A\&A,  301, 757

\bibitem[Bonazzola \& Gourgoulhon(1994)]{Bonazzola:1994}
  Bonazzola,~S. \& Gourgoulhon,~E. 
  1994, CQG, 11, 1775

\bibitem[Bonazzola \& Gourgoulhon(1996)]{Bonazzola:1995rb}
  Bonazzola,~S., \& Gourgoulhon,~E. 
  1996, A\&A  312, 675

\bibitem[Bonazzola et al.(1998)]{bona} Bonazzola, S., 
Gourgoulhon, E., \& Marck, J.-A.\ 1998, \prd, 58, 104020 

\bibitem[Braithwaite \& Spruit(2004)]{Braithwaite:2005ps}
  Braithwaite,~J. and Spruit,~H.~C.,
  2004, Nature 431, 819

\bibitem[\protect\citeauthoryear{Broderick et al.}{2000}]{brod} 
  Broderick, A., Prakash, M., \& Lattimer, J.~M.\ 2000, ApJ, 537, 351 

\bibitem[Cardall et al.(2001)]{Cardall:2001}
  Cardall,~C.~Y., Prakash,~M., \& Lattimer,~J.~M. 
  2001, ApJ, 554, 322 

\bibitem[Carter(1969)]{Carter:1969}
  Carter,~B. 
  1969, J.\ Math.\ Phys.\ 10, 70

\bibitem[Chandrasekhar \& Fermi(1953)]{Chandra:1953}
  Chandrasekhar,~S., \& Fermi,~E. 
  1953, ApJ  118, 116

\bibitem[Cook et al.(1992)]{Cook:1992}
  Cook,~G.~B., Shapiro,S.~L. \& Teukolsky,~S.~A., 
  1992, ApJ 398, 203, 1994, APJ 422, 227

\bibitem[Cutler(2002)]{Cutler:2002nw}
  Cutler,~C. 2002, 
  PRD 66, 084025

\bibitem[Dessart et al.(2007)]{luc2007} 
  Dessart, L., Burrows, A., Livne, E., \& Ott, C.~D.\ 2007, 
  APJ, 669, 585 

\bibitem[Douchin \& Haensel(2001)]{douchin}
  Douchin, F., \& Haensel, P.\ 2001, A\&A, 380, 151 

\bibitem[Ferrario \& Wickramasinghe(2006)]{Ferrario:2007bt}
  Ferrario,~L. \& Wickramasinghe,~D. 2006, 
  MNRAS, 367, 1323

\bibitem[Friedman \& Pandharipande(1981)]{fp81}
  Friedman, B., \& Pandharipande, V.~R.\ 1981, Nuclear Physics A, 361, 502 

\bibitem[Friedman et al.(1988)]{Friedman:1988}
  Friedman, J.~L., Ipser, J.~R., \& Sorkin, R.~D., 1988, Apj 325, 722

\bibitem[Geppert \& Rheinhardt(2006)]{Geppert:2006cp}
  Geppert,~U. \& Rheinhardt,~M.
  2006, A\& A, 456, 639

\bibitem[Glendenning(2001)]{gled}
  Glendenning, N.~K.\ 2001, 
  Physics. Rep., 342, 393 

\bibitem[Goldreich \& Julian(1969)]{Goldreich:1969}
  Goldreich,~P. \& Julian,~W.,
  1969, APJ, 238, 991

\bibitem[Goldreich \& Reisenegger(1992)]{Goldreich}
  Goldreich,~P. \& Reisenegger,~A.,
  1992, APJ, 395, 250

\bibitem[Gourgoulhon \& Bonazzola(1993)]{Gourgoulhon:1993}
  Gourgoulhon,~E., \& Bonazzola,~S. 1993, 
  PRD, 48, 2635 

\bibitem[Gourgoulhon \& Bonazzola(1994)]{Gourgoulhon:1994}
  Gourgoulhon,~E., \& Bonazzola,~S. 1994
  CQG,  11, 443

\bibitem[Harding \& Lai(2006)]{Harding:2006} 
  Harding, A.~K., \& Lai, D.\ 2006, 
  Reports of Progress in Physics, 69, 2631 

\bibitem[Heger et al.(2005)]{Heger:2004qp}
  Heger,~A., Woosley,~S.~E., and Spruit,~H.~C.,
  2005, APJ  626, 350

\bibitem[Hachisu(1986)]{Hachisu:1986}
  Hachisu, I. 1986, 
  ApJS, 61, 479 

\bibitem[Ioka \& Sasaki(2003)]{Ioka:2003}
  Ioka,~K. \& Sasaki,~M. 2003,
  PRD  67, 124026

\bibitem[Ioka \& Sasaki(2004)]{Ioka:2004}
  Ioka,~K. \& Sasaki,~M. 2004,
  ApJ   600, 296

\bibitem[Kiuchi \& Kotake(2008)]{Kiuchi:2007pa}
  Kiuchi,~K., \& Kotake,~K. 
  2008, MNRAS 385, 1327

\bibitem[Kiuchi \& Yoshida(2008)]{Kiuchi:2008ch}
  Kiuchi,~K. \& Yoshida,~S. 
  2008, PRD 78, 044045

\bibitem[Kiuchi et al.(2008)]{Kiuchi:2008ss}
  Kiuchi,~K., Shibata,~M. \& Yoshida,~S., 
  2008, PRD 78, 024029

\bibitem[Komatsu et al.(1989)]{Komatsu:1989}
  Komatsu,~H., Eriguchi,~Y., \& Hachisu,~I., 
  MNRAS, 237, 355 (1989), 239, 153 (1989)

\bibitem[Konno et al.(1999)]{Konno:1999zv}
  Konno,~K., Obata,~T.\& Kojima,~Y. 
  1999, A\&A, 352, 211

\bibitem[Kotake et al.(2004)]{Kotake:2004} 
  Kotake, K., Sawai, H., Yamada, S., \& Sato, K.\ 2004, ApJ, 608, 391 

\bibitem[Kotake et al.(2006)]{Kotake:2006} 
  Kotake, K., Sato, K., \& Takahashi, K.\ 2006, 
  Reports of Progress in Physics, 69, 971 

\bibitem[Kouveliotou(1998)]{kouve} 
  Kouveliotou, C., et al.\ 1998, Nature, 393, 235 

\bibitem[Lattimer \& Prakash(2007)]{Lattimer:2006}
  Lattimer,~J.~M. \& Prakash,~M. 2007 
  Phys.\ Rept.\  442, 109

\bibitem[Lattimer \& Swesty(1991)]{LS}
  Lattimer, J.~M., \& Douglas Swesty, F.\ 1991, Nuclear Physics A, 535, 331 

\bibitem[Livne et al.(2007)]{Livne:2007} 
  Livne, E., Dessart, L., Burrows, A., \& Meakin, C.~A.\ 2007, 
  ApJS, 170, 187 

\bibitem[Lyne \& Graham-Smith(2005)]{Lyne} 
  Lyne A. \& Graham-Smith, F. {\em Pulsar Astronomy} 
  (Cambridge University Press, 2005). 

\bibitem[Miketinac(1973)]{Miketinac:1973}
  Miketinac,~M.~J. 1973
  Ap\&SS, 22, 413

\bibitem[Morrison et al.(2004)]{morrison} 
  Morrison, I.~A., Baumgarte, T.~W., \& Shapiro, S.~L.\ 2004, ApJ, 610, 941 

\bibitem[Nozawa et al.(1998)]{nozawa}
  Nozawa, T., Stergioulas, N., Gourgoulhon, E., \& Eriguchi, Y.\ 1998, 
  A\&A Sup. Ser., 132, 431 

\bibitem[Obergaulinger et al.(2006)]{Obergaulinger:2006}
  Obergaulinger, M., Aloy, M.~A., M\"{u}ller, E.\ 2006, 
  A\&A, 450, 1107 

\bibitem[Oron(2002)]{Oron: 2002}
  Oron,~A., 2002, 
  PRD, 66, 023006

\bibitem[Pons 
\& Geppert(2007)]{pons} Pons, J.~A., \& Geppert, U.\ 2007, \aap, 470, 303 

\bibitem[Pandharipande \& Ravenhall(1989)]{pand}
  Pandharipande, V.~R., \& Ravenhall, D.~G.\ 1989, NATO ASIB Proc.~205: 
  Nuclear Matter \& Heavy Ion Collisions, 103 

\bibitem[Reisenegger(2001)]{Reisenegger2001} Reisenegger, A.\ 2001, 
\apj, 550, 860 

\bibitem[Reisenegger(2008)]{Reisenegger:2008yk}
  Reisenegger,~A.,
  arXiv:0809.0361 [astro-ph]

\bibitem[Sawai et al.(2008)]{sawai2008} 
  Sawai, H., Kotake, K., \& Yamada, S.\ 2008, 
  APJ, 672, 465 

\bibitem[Shapiro \& Teukolsky(1983)]{shapiro}
  Shapiro, S.~L., \& Teukolsky, S.~A.\ 1983, 
  Research supported by the National Science Foundation.~New York, Wiley-Interscience, 1983, 645 p.,

\bibitem[Shapiro et al.(1990)]{Shapiro:1990}
  Shapiro,~S.~L., Teukolsky,~S.A., \& Nakamura,~T. 1990, 
  ApJ, 357, L17 

\bibitem[Shen et al.(1998a)]{shen98}
  Shen, H., Toki, H., Oyamatsu, K., Sumiyoshi, K.  1998a, 
  Nuclear Physics, A637, 435, 109, 301

\bibitem[Shen et al.(1998b)]{shenprog}
  Shen, H., Toki, H., Oyamatsu, K., \& Sumiyoshi, K.\ 1998b, 
  Progress of Theoretical Physics, 100, 1013 

\bibitem[Shibata et al.(2005)]{Shibata:2005ss}
  Shibata,~M., Taniguchi,~K., \& Uryu,~K., 2005, 
  PRD 71, 084021

\bibitem[Shibata et al.(2006)]{Shibata:2006hr}
  Shibata,~M., Liu,~Y.~T., Shapiro,~S.~L., \& Stephens,~B.~C. 2006
  PRD 74, 104026

\bibitem[Spruit(2007)]{Spruit}
  Spruit,~H.~C. astro-ph/0711.3650. 

\bibitem[Sumiyoshi et al.(2005)]{sumi}
  Sumiyoshi, K., Yamada, S., Suzuki, H., Shen, H., Chiba, S., \& Toki, H.\ 
  2005, ApJ, 629, 922 

\bibitem[Takiwaki et al.(2004)]{takiwaki1} Takiwaki, T., Kotake, 
K., Nagataki, S., \& Sato, K.\ 2004, \apj, 616, 1086 

\bibitem[Takiwaki et al.(2009)]{takiwaki} Takiwaki, T., Kotake, 
K., \& Sato, K.\ 2009, \apj, 691, 1360 

\bibitem[Tayler(1973)]{Tayler: 1973}
  Tayler,~R.~J., 1973
  MNRAS, 161, 365

\bibitem[Thompson \& Duncan(1993)]{Thompson:1993hn}
  Thompson,~C. \& Duncan,~R.~C. 1993,
  ApJ, 408, 194

\bibitem[Thompson \& Duncan(1995)]{Thompson:1995gw}
  Thompson,~C. \& Duncan,~R.~C. 1995,
  MNRAS,  275, 255

\bibitem[Thompson \& Duncan(1996)]{Thompson:1996pe}
  Thompson,~C. \& Duncan,~R.~C. 1996,
  ApJ 473, 322

\bibitem[Tomimura \& Eriguchi(2005)]{Tomimura:2005}
  Tomiumra,~Y. \& Eriguchi,~Y. 2005 
  MNRAS,  359, 1117

\bibitem[Trehan \& Uberoi(1972)]{trehan1972}
  Trehan,~S.~K. \& Uberoi,~M.~S. 1972
  ApJ, 175, 161 

\bibitem[Wald(1984)]{Wald:1984}
  Wald,~R.~M. 1984, 
  {\it General Relativity} (The University of Chicago Press),

\bibitem[Watts(2006)]{anna}
  Watts, A.\ 2006, 36th COSPAR Scientific Assembly, 36, 168 

\bibitem[Wiringa et al.(1988)]{wiringa}
  Wiringa, R.~B., Fiks, V., \& Fabrocini, A.\ 1988, PRC, 38, 1010 

\bibitem[Woods \& Thompson(2004)]{wod} 
  Woods,~P.~M. \& Thompson,~C. 2004,
  arXiv:astro-ph/0406133.

\bibitem[Yoshida \& Eriguchi(2006)]{Yoshida:2006a}
  Yoshida,~S. \& Eriguchi,~Y. 2006,
  ApJS,  164, 156

\bibitem[Yoshida et al.(2006)]{Yoshida:2006b}
  Yoshida,~S., Yoshida,~S., \& Eriguchi,~Y. 2006,
  ApJ,  651, 462


\end{thebibliography}
\end{document}